\documentclass[prodmode,acmtois]{acmsmall}

\usepackage{booktabs}

\usepackage{amsmath}  % \begin{multiline}, \begin{gathered}
\usepackage{accents}  % \undertilde

\usepackage{siunitx}
\usepackage{multirow}

\usepackage{bigstrut}

\usepackage{url}

\newcommand{\defineterm}[1]{\emph{#1}}

\newcommand{\metname}[1]{\ensuremath{\mathrm{#1}}}
\newcommand{\recall}{\metname{Rec}}
\newcommand{\precision}{\metname{Prec}}

\newcommand{\est}[1]{\widehat{#1}}
\newcommand{\var}{\ensuremath{\mathrm{Var}}}
\newcommand{\cov}{\ensuremath{\mathrm{Cov}}}

\newcommand{\expected}[1]{\ensuremath{\mathbb{E}}[#1]}

\newcommand{\lowerbound}[1]{\underline{#1}}
\newcommand{\upperbound}[1]{\overline{#1}}

\newcommand{\dif}{\mathrm{d}}

\newcommand{\probdist}[1]{\ensuremath{\mathit{#1}}}
\newcommand{\betadist}{\probdist{Beta}}
\newcommand{\betabindist}{\probdist{BetaBin}}
\newcommand{\hypergeomdist}{\probdist{HyperGeom}}
\newcommand{\unifdist}{\probdist{Unif}}

\DeclareMathOperator{\logit}{logit}

\newcommand{\given}{\,|\,}

\usepackage[normalem]{ulem}     % Part of the standard distribution
\usepackage[xcolor]{changebar}
\cbcolor{gray}

{\end{changebar}\cbcolor{gray}}

\title{Approximate Recall Confidence Intervals}

\author{WILLIAM WEBBER \affil{The University of Maryland\footnote{Work
performed in part while author was at the University of Melbourne.}}}

\begin{abstract}
Recall, the proportion of relevant documents retrieved, is an 
important measure of effectiveness in information 
retrieval, particularly in the legal, 
patent, and medical domains.  Where document sets are too large
for exhaustive relevance assessment, recall can be estimated by assessing a
random sample of documents; but
an indication of the reliability of this estimate is also required.
In this article, we examine several methods for estimating 
two-tailed recall confidence intervals.  We
find that the normal approximation in current use provides 
poor coverage in many circumstances, even when adjusted 
to correct its inappropriate symmetry.
Analytic and Bayesian methods based on the ratio of binomials
are generally more accurate, but are inaccurate on small 
populations.  The method we recommend derives beta-binomial
posteriors on retrieved and unretrieved yield, with fixed hyperparameters,
and a Monte Carlo estimate of the posterior distribution of recall.
We demonstrate that this method gives mean coverage at or near
the nominal level, across several scenarios, while being balanced
and stable.  We offer advice on sampling design, including the
allocation of assessments to the retrieved and unretrieved segments,
and compare the proposed beta-binomial with the officially
reported normal intervals for recent TREC Legal Track iterations.
\end{abstract}

\category{G.3}{Mathematics of Computing}{Probability and Statistics}[experimental design]
\category{G.3}{Mathematics of Computing}{Probability and Statistics}[distribution functions]
\category{H.3.4}{Information Storage and Retrieval}{Systems and Software}[performance evaluation (efficiency and effectiveness)]

\terms{Measurements, Experimentation, Verification}

\keywords{Posterior distributions, probabilistic models}

\acmformat{Webber, W., Approximate recall confidence intervals}

\begin{document}

\bibliographystyle{acmsmall}

\begin{bottomstuff}
This work was supported in part by the Australian Research Council.
This material is based upon work supported by the National Science
Foundation under Grant No. 1065250.  Any opinions, findings, and
conclusions or recommendations expressed in this material are those of
the authors and do not necessarily reflect the views of the National
Science Foundation.

Author's address:
College of Information Studies,
University of Maryland,
College Park, MD 20742, The United States of America.  
Email: wew@umd.edu.
\end{bottomstuff}

\maketitle

\section{Introduction}

There are many ways to measure the effectiveness of the retrieval
of a set of documents from a document corpus. 
One of these is recall, the proportion of the 
relevant documents in the corpus that are retrieved.  
High recall is important for domains that require comprehensive
retrieval,  including patent search, medical
literature reviews, and document discovery for civil
litigation (or \defineterm{e-discovery}).  
To calculate recall, however, we must know
the number of relevant documents, or \defineterm{yield}, of the corpus.
Most corpora are too large for exhaustive relevance assessment; in
some cases, even the retrieved documents are too numerous.
Retrievals in the Legal Interactive Task of TREC can be in the
tens or even hundreds of thousands~\cite{legal08:trec}, and
document productions in practicing e-discovery can readily
run to the hundreds of thousands, extracted from corpora in
the millions of documents~\cite{rko10:jasist}.

Where exhaustive assessment is impractical,
the yields of the retrieved and unretrieved segments of
the corpus can instead be estimated, by drawing a random sample of
documents from each segment,
and assessing sampled documents for relevance.
The proportion of relevant documents, 
or \defineterm{prevalence}, in the sample gives
a point estimate of prevalence in the segment, and hence
of yield and recall (Section~\ref{sect:est-rec}).  Retrieved
and unretrieved segments can be sampled at different rates, and
each segment may be stratified, taking advantage of auxiliary
evidence about stratum prevalence to improve estimator
accuracy (Section~\ref{sect:strat}).

Point estimates, however, are subject to sampling error;
random variability in the proportion of relevant documents
sampled in each segment may lead them to under- or overestimate
recall.  Errors in the point estimate
can cause incorrect decisions being made; for instance, a
production in e-discovery may be accepted as sufficiently thorough
when it is in fact seriously incomplete.  Moreover, from the
point estimate alone, one cannot tell how likely these erroneous
results are to occur.  Larger samples lead to
more accurate estimates, but this is not expressed in the point
estimate, removing much of the incentive for more thorough
evaluation.  

Errors in point estimates are particularly serious where segments
are large and prevalence is low.  Such a condition frequently
applies to the unretrieved segment of the corpus, since the retrieval
will only return a fraction of the corpus but will (if effective) 
have concentrated relevant documents in this fraction,
leaving them sparse in the larger remnant.  Small samples may
find no relevant documents in the unretrieved segment,
leading to the misleading impression of perfect recall.  Indeed,
when retrieval implementors evaluate their own work, using
the point estimate can create a perverse incentive to reduce
sample size, so as to minimize the risk of finding an unretrieved 
relevant document.

Instead of relying on point estimates alone,
an interval may be calculated within which recall falls, with $1 - \alpha$
confidence.  A \defineterm{two-tailed confidence interval} provides
a finite upper and lower bound, while a one-tailed interval give
a lower or an upper bound alone; this paper examines two-tailed intervals.
An \defineterm{exact} confidence interval, constructed
from the sampling distribution of the statistic, guarantees that
for any true parameter, at least $1 - \alpha$ of resampled and recalculated
intervals would cover that parameter.  
Where exact intervals are unavailable, or too complex to compute or analyze,
\defineterm{approximate} intervals may be used instead~\cite{smithson02:ci}.
In addition, exact intervals often \defineterm{over-cover}
(provide mean coverage greater than $1 - \alpha$)
discrete statistics (Section~\ref{sect:exact-ci}), whereas 
approximate intervals may give nominal mean coverage 
(Section~\ref{sect:approx-ci}).  
This article considers approximate intervals.

We examine several 
confidence intervals on recall (Section~\ref{sect:rec-ci}).  
The first treats recall as a binomial
proportion on relevant documents
(Section~\ref{sect:naive-binom})~\cite{ssm91:jclep}, but
assumes the set of relevant documents is sampled at an equal rate.
A second method, previously employed in e-discovery~\cite{legal08:trec}, 
uses a normal approximation, 
a maximum-likelihood estimate (MLE) of variance, and the propagation
of error (Section~\ref{sect:normal-mle}); but this method often under-covers
recall.  The remaining methods are new.

For extreme prevalence, the Normal MLE interval underestimates
uncertainty, and its assumption of symmetry is incorrect.  
Adjusting the normal approximation, by adding one or two 
to the positive and negative sample counts, 
gives a superior binomial confidence 
interval~\cite{ac98:tas,greenland01:tas}.  We
apply a similar adjustment to the normal 
approximation of the recall interval 
(Section~\ref{sect:normal-adj}).  The adjustment
improves coverage in some circumstances, 
but has mixed overall accuracy.

The normal intervals assume that recall is approximately
normal in its sampling distribution; but this is not in general
the case (Section~\ref{sect:rec-dist}).  A closer
approximation models sampled recall as a function of
the ratio between two binomial variables.  We adapt
an interval on a binomial ratio \cite{koopman84:biom} 
to bound recall in Section~\ref{sect:koopman}.
The method is generally accurate, but
has no finite population adjustment, and so
overstates the interval where sample size is a
substantial proportion of a segment.

The above intervals are derived from the sampling 
distribution of recall or its components, such as yield.  
An alternative, Bayesian approach is to infer a 
distribution over recall (Section~\ref{sect:bayes}), 
and take the $\alpha / 2$ and $1 - \alpha / 2$ quantiles as the interval.
We infer distributions over retrieved and unretrieved
yield, and generate from these a Monte Carlo estimate 
of the recall quantiles \cite{buck84:biom,cs99:jcgs}.

A beta posterior with the Jeffreys prior
gives accurate intervals on the binomial
proportion~\cite{bcd01:statsci} (Section~\ref{sect:approx-ci}),
and beta posteriors on retrieved and unretrieved yield
generate mostly accurate intervals on recall
(Section~\ref{sect:jeffreys}).
However, the beta interval become less accurate as sample
proportion grows, even with a finite population adjustment,
as it omits the dependence between what is seen in the sample and what
remains in the population.

Sampling without replacement from a finite binomial
population is most precisely modelled by a hypergeometric
distribution.  The conjugate distribution (Section~\ref{sect:bayes})
to the hypergeometric is the beta-binomial.  Our final 
approach derives beta-binomial posteriors
on retrieved and unretrieved yield
(Section~\ref{sect:hypergeom}), and is the 
most accurate of those considered.  The uniform and the
information-theoretic
most conservative prior (Section~\ref{sect:beta-bin-prior})
are slightly, and differently, unbalanced; a
simple prior that sets the beta-binomial hyperparameters to
$0.5$ (the \defineterm{half prior}) gives best results.

The recall interval methods are evaluated in Section~\ref{sect:eval}
against four criteria:  first, mean coverage should
be close to $1 - \alpha$; second, the standard error of the coverage 
around this point should be low; 
third, an equal proportion of uncovered parameters should
fall below as above the interval; and fourth, interval width
should be the minimum required to achieve accurate mean coverage.
These criteria must be evaluated over
true parameter distributions, and we define
three such distributions or scenarios (Section~\ref{sect:eval-meth}): a broad
neutral one; a second that emulates an e-discovery environment; 
and a third that explores the finite population case of
a small population and large sample.
The beta-binomial with half prior is found to be the most accurate
method on all three scenarios.

Section~\ref{sect:design} examines experimental design,
including the allocation of assessments to retrieved and unretrieved
segments, and the effect of sample size on interval width.  In
Section~\ref{sect:trec-int}, we calculate beta-binomial intervals
on TREC Legal Interactive participants, and compare them with the officially
reported, normal approximation intervals.  Finally, Section~\ref{sect:concl}
summarizes our findings, and lays out future work.

\section{Preliminaries}
\label{sect:prelim}

This section sets out preliminary materials, used in
Section~\ref{sect:rec-ci} to design recall confidence
intervals.  We begin by deriving a point estimate
of recall (Section~\ref{sect:est-rec}), then extend
it to stratified sampling (Section~\ref{sect:strat}).  
The hypergeometric, binomial, and normal distributions, used to
construct recall intervals, are
introduced in Section~\ref{sect:dist}.  
The point estimate of recall is found in Section~\ref{sect:rec-dist}
to be strongly biased, and its sampling distribution highly non-normal,
for a representative scenario.
Section~\ref{sect:exact-ci}
defines exact confidence intervals, while Section~\ref{sect:approx-ci}
describes approximate intervals, and why they can be
preferable to exact ones.  
Bayesian prior and posterior
distributions are introduced in Section~\ref{sect:bayes}, and
Section~\ref{sect:beta-bin-prior}
describes the beta-binomial distribution, which is conjugate prior 
to the hypergeometric distribution.
Interval estimation through Monte Carlo simulation
is examined in Section~\ref{sect:mc-int}.
Finally, Section~\ref{sect:prop-err} explains the
propagation of error, used for normal intervals,
and derives a propagation of error expression for recall.
(The reader may wish to skim this section on first reading,
and refer back to it to clarify components of the intervals 
described in Section~\ref{sect:rec-ci}.)

\begin{table}
\tbl{Notation. \label{tbl:notation}}{
\begin{tabular}{cp{0.6\textwidth}l}
\toprule
Symbol  &  Meaning  & Notes \\
\midrule
$N_*$     &  Size of corpus \\
$N_1$   &  Size of retrieved segment \\
$N_0$   &  Size of unretrieved segment & $N_* = N_1 + N_0$ \\
$R_*$, $R_1$, $R_0$ & Number of relevant documents (yield) in above
    populations & $R_* = R_1 + R_0$ \\
$\pi_*$, $\pi_1$, $\pi_2$ & Prevalence of relevant documents in population 
        & $\pi = R_* / N_*$ \\
$n_*$, $n_1$, $n_0$ & Size of samples & $n_* = n_0 + n_1$\\
$r_*$, $r_1$, $r_0$ & Number of relevant documents (yield) in sample
        & $r_* = r_1 + r_0$ \\
$p_*$, $p_1$, $p_2$ & Prevalence of relevant documents in sample
        & $p_* = r_* / n_*$ \\
$N$, $n$, $R$, $\ldots$ & Population size, sample size etc.\ 
    for unspecified segments \\
$N_s$, $n_s$, $R_s$, $\ldots$ & Population size, sample size etc.\ for
    stratum $s$ (Section~\ref{sect:strat}) \\
\bottomrule
\end{tabular}
}
\end{table}

\subsection{Estimating recall}
\label{sect:est-rec}

\newcommand{\mco}[1]{\multicolumn{1}{c}{#1}}
\newcommand{\cn}[1]{\centering\raisebox{-1em}[1em][2em]{#1}}

\begin{table}[t!]
\tbl{Document counts in a retrieval. \label{tbl:doc-counts}}{
\begin{tabular}{p{2cm} | p{2cm} | p{2cm} | c p{2cm} }
\mco{} & \mco{\cn{Relevant}} & \mco{\cn{Not relevant}} & \mco{} \\
\cline{2-3}
\cn{Retrieved} & \cn{$R_1$} & \cn{$N_1 - R_1$} && {\cn{$N_1$}} \\
\cline{2-3}
\cn{Not retrieved} & \cn{$R_0$} & \cn{$N_0 - R_0$} && {\cn{$N_0$}} \\
\cline{2-3}
\mco{}         & \mco{\cn{$R_*$}} 
               & \mco{\cn{$N_* - R_*$}} && {\cn{$N_*$}} \\
\end{tabular}
}
\end{table}

A set retrieval returns from a corpus the set of documents that it
considers relevant.  Such a retrieval
is equivalent to a binary classification of the corpus documents
into relevant and irrelevant classes.  Let
the number of documents in the corpus be $N_*$, the number in the
retrieved segment be $N_1$, and the number in the unretrieved segment
be $N_0 = N_* - N_1$. Some $R_*$ of documents in the corpus 
are actually relevant to the topic; $R_1$ of these fall in the retrieved
segment, and $R_0$ in the unretrieved segment. 
Our notation is laid in Table~\ref{tbl:notation}, and the document
counts are summarized in Table~\ref{tbl:doc-counts}.
We refer to the
number of relevant documents in a set as the set's \defineterm{yield}.
The recall of a
retrieval is the proportion of relevant documents retrieved:
\begin{equation}
\recall = \frac{R_1}{R_*} = \frac{R_1}{R_1 + R_0} 
\label{eqn:rec}
\end{equation}

Corpus yield is rarely known in advance, and the corpus is 
generally too large to determine it by exhaustive assessment;
even the retrieval may be too large to be exhaustively assessed.
Instead, a sample-based estimate may be made.
A size-$n$ (without-replacement) simple random sample 
from a population of $N$ elements 
is one in which each subset of $n$ items has the same probability
$1 / {N \choose n}$ of being the sample, and each
element therefore has $n/N$ probability of being included 
in the sample~\cite{ssw92:mass}.
To estimate recall, we draw a simple random sample of documents
from the retrieved and unretrieved segments; assess these documents
for relevance; and use the assessed sample to estimate yield
in each segment.  Let $n_1$ and $n_0$ be the number sampled from the retrieved
and unretrieved segments, and $r_1$ and $r_0$ be the number
relevant in these samples.  The prevalence of relevant documents in the
retrieved sample, $p_1$, is an estimator of 
prevalence in the retrieved population, ${\pi}_1$:
\begin{equation*}
p_1 = \est{\pi}_1 = r_1 / n_1 \; ,
\end{equation*}
and likewise for the unretrieved sample prevalence, $p_0$, on the
unretrieved segment.  Then, estimators for the yields of the each
segment are:
\begin{equation*}
\est{R}_1 = N_1 \est{\pi}_1 \quad ; \quad \est{R}_0 = N_0 \est{\pi}_0 \; ;
\end{equation*}
and an estimator for recall is:
\begin{equation}
\est{\recall} = \frac{\est{R}_1}{\est{R}_1 + \est{R}_0} 
              = \frac{\est{R}_1}{\est{R}_*}\; .
\label{eqn:est-rec}
\end{equation}
There is an obvious dependence between the numerating
sample variable $\est{R}_1$ and the denominating 
sample variable $\est{R}_* = \est{R_1} + \est{R_0}$,
since the latter includes the former; such dependence
makes estimation of variance more complicated 
(Section~\ref{sect:prop-err}).
The dependence can be avoided by rewriting the recall
estimator as:
\begin{equation}
\est{\recall} = \frac{1}{1 + \est{R}_0/\est{R}_1} \; .
\label{eqn:est-rec-alt}
\end{equation}
As an estimation of a function of a ratio, the recall estimator 
is biased; that is, $\expected{\est{\recall}} \neq \recall$, 
the expectation of the estimator does not equal
true recall~\cite{hr54:nature,cochran77}.  
The bias is negligible for mediate prevalences and large samples, 
but can be serious for extreme prevalences and small
sample sizes (Section~\ref{sect:rec-dist}).

\subsection{Stratified sampling}
\label{sect:strat}

The retrieved and the unretrieved segments can be subdivided
into disjoint strata, from each of which a simple random sample
is drawn.  Estimate accuracy is improved by stratification if
prevalence differs between strata, and further still if
more samples are allocated to strata that have mediate prevalence,
and therefore higher sample variance~\cite{tho02}. 
If several set retrievals are being evaluated over 
the one corpus, the
intersections between them form natural strata
\cite{legal07:trec}.  

The estimated yield $R_s$ of stratum $s$
is derived from the size of the stratum $N_s$ and
the proportion of the stratum sample $r_s / n_s$ that is
relevant:
\begin{equation}
\est{R}_s = N_s r_s / n_s \; .
\end{equation}
The estimated yield of segment $T$ is the sum
of the estimated yields of the strata in $T$:
\begin{equation}
\est{R}_T = \sum_{s \in T} \est{R}_s \; .
\label{eqn:sum-strata}
\end{equation}
Recall is estimated using these segment estimates in
Equation~\ref{eqn:est-rec}.

\subsection{Hypergeometric, binomial, and normal distributions}
\label{sect:dist}

A segment or stratum can be viewed as a binomial population 
of relevant and irrelevant
documents.  If we draw a simple random sample of $n$ documents
from a population of size $N$, in which $R$ documents are 
relevant, then the probability that the number $r$ of relevant documents
in the sample takes on a particular value $k$ is given by the
hypergeometric distribution:
\begin{equation}
\Pr(r = k \given N, R, n) 
  \; = \; f(k) 
  \; = \; \frac{{R \choose k}{N - R \choose n - k}}{{N \choose n}} 
\; ,
\end{equation}
where ${a \choose b} = a! / (b! (a - b)!)$.  The hypergeometric
distribution is so named because its successive terms in $k$ follow
a hypergeometric series, in which 
$f(k + 1) / f(k) = g(k)$, where $g(\cdot)$ is some ``simple'' 
function of $k$.  The prefix ``hyper'' comes from viewing this
as an extension of the geometric series, where $g( \cdot )$ is
a constant.  Setting $g(k) = k$ and $f(0) = 1$ gives the
exponential series.  For the hypergeometric distribution, we have:
\begin{equation}
g(k) = \frac{(R - k) (n - k)}{(k + 1) (N - R - n + k + 1)}
\end{equation} 
See \citeN{dutka84:ahes} for an historical perspective.

As population size $N$ increases, the change that each
draw makes upon the original population prevalence 
$\pi = N/R$ diminishes, and the probability of 
sampling $k$ relevant documents is 
approximated with increasing closeness by the binomial distribution:
\begin{equation}
\Pr(r = k \given \pi, n) = {n \choose k} \pi ^ k (1 - \pi)^{n - k} \; .
\end{equation}
The approximation
is weaker for smaller populations $N$, larger samples $n$, and
prevalences $\pi$ more divergent from $0.5$.
For large $n$ and $\pi$ not close to $0$ or $1$, the binomial distribution
is in turn approximated by the normal distribution:
\begin{equation}
\Pr(r = k \given \pi, n) = \phi(k; n\pi, n\pi(1 - \pi) ) \; ,
\end{equation}
where $\phi( \cdot ; \mu, \sigma^2)$ is the normal distribution function
with mean $\mu$ and variance $\sigma^2$.

\begin{figure}
\centering
\includegraphics{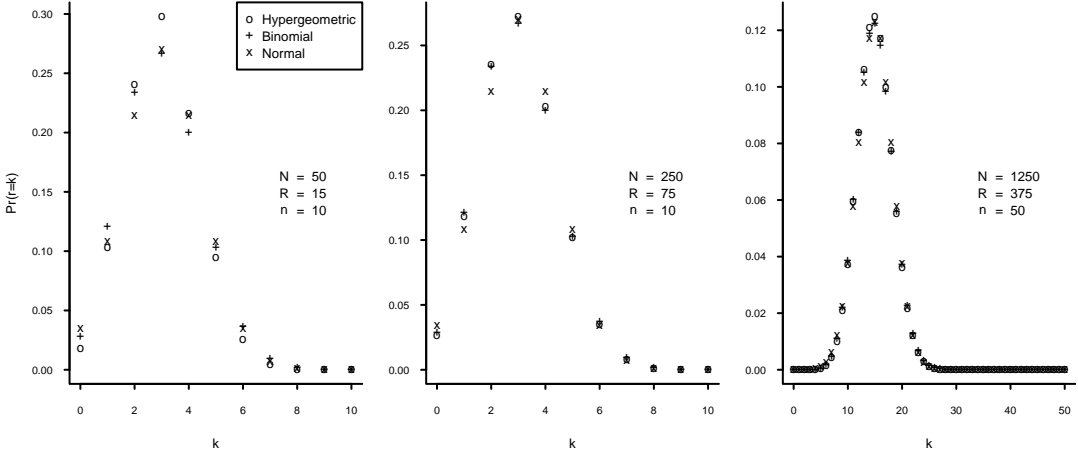}
\caption{Probability of drawing $k$ relevant documents in simple
random sample of size $n$ from a population with $N$ documents, $R$
of them relevant, as calculated by the hypergeometric distribution
and by the binomial and normal approximations,
for different $n$, $N$, and $R$.  The normal approximation uses
continuity correction; that is, $\Pr(r=k) = \Pr(r < k + 0.5) -
\Pr(r < k - 0.5)$.}
\label{fig:hyp-bin-norm}
\end{figure}

Figure~\ref{fig:hyp-bin-norm} compares 
the hypergeometric, binomial, and discretized normal distributions,
for the one population prevalence $R/N = \pi = 0.3$
and different population and
sample sizes, $N$ and $n$.  The hypergeometric is the correct
without-replacement sampling distribution.
Where $n$ is not small relative to $N$ (left), the binomial 
and normal approximations 
overstate the probability of sample prevalence $p$ diverging from 
population prevalence $\pi$.  For small $n$ and $N \gg n$ 
(middle), the binomial approximation is close to the hypergeometric,
but the (incorrectly symmetric) normal approximation overstates the probability
of sample prevalence being towards $0.5$.  Increasing $n$, 
while holding $n/N$ fixed (right), brings the normal
approximation closer to the hypergeometric \cite{feller45:ams,nich56:ams}.

\subsection{The sampling distribution and bias of the recall estimator}
\label{sect:rec-dist}

\begin{table}
\tbl{Example scenario used in Figure~\ref{fig:rec-smpl-dist-bar}.  
\label{tbl:bias-rec-eg}
}{
\begin{tabular}{l c r  r  r}
\toprule
Segment     && Yield      & Population size & Sample size \\
\midrule
Retrieved   &&  $1{,}000$ &   $2{,}000$ & $100$ \\
Unretrieved &&  $3{,}000$ & $100{,}000$ & $100$ \\
\midrule
Total       &&  $4{,}000$ & $102{,}000$ & $200$ \\
\bottomrule
\end{tabular}}
\end{table}

\begin{figure}
\centering
\includegraphics{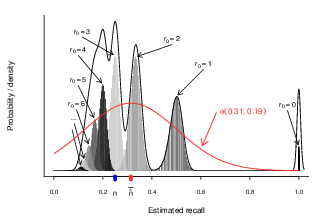}
\caption{Sampling distribution of recall for example
scenario (Table~\ref{tbl:bias-rec-eg}).  The bars show 
sampling distributions induced by $r_1$ over different
values of $r_0$.  The full distribution is approximated
by a Gaussian kernel density estimate.  
If $r_0 = 0$, estimated recall is precisely $1$;
the density here is for comparison of area only.  True
recall, $R$, and estimator mean,
$\bar{\est{R}}$, are marked; the estimator has a
strong positive bias.  A normal curve with the mean and
standard deviation of the sampling distribution is also shown.
}
\label{fig:rec-smpl-dist-bar}
\end{figure}

Deriving a closed-form expression for the sampling distribution of
recall is not straightforward; but the distribution for
any particular population and sample size can be calculated
by brute force.  Figure~\ref{fig:rec-smpl-dist-bar}
shows the sampling distribution of the recall estimator 
(Equation~\ref{eqn:est-rec})
for a representative scenario,
given in Table~\ref{tbl:bias-rec-eg}.
The small number of discrete values that $r_0$ can take on, 
combined with estimator's skew, produces a complex, skewed,
multi-modal distribution, having wide gaps
around high recall estimates.  The
normal distribution, also shown in Figure~\ref{fig:rec-smpl-dist-bar}, 
is a poor approximation.

\begin{figure}
\centering
\includegraphics{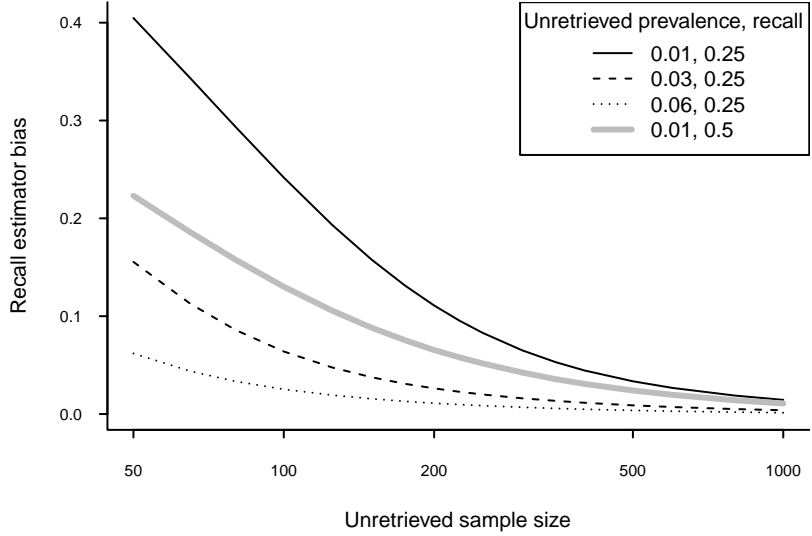}
\caption{Bias of recall estimator for scenario described in
Table~\ref{tbl:bias-rec-eg} but with varying unretrieved
sample sizes ($x$ axis) and unretrieved yield (and therefore
prevalence).  Recall and precision are
held fixed in the first three cases by scaling retrieved 
population size and yield by the same amount as unretrieved yield;
in the fourth case, retrieved yield and population size is held fixed,
and recall increases with the decrease in unretrieved yield.
Note the log scale on the $x$ axis.}
\label{fig:rec-bias}
\end{figure}

The recall estimator also has a strong positive 
bias for this scenario: true recall is $0.25$, but the
mean of the estimator is $0.31$.  The bias is the result of
the estimator's skew, clearly visible in
Figure~\ref{fig:rec-smpl-dist-bar}; this is one occasion in which
the bias in the estimation of a ratio is non-negligible.  
Figure~\ref{fig:rec-bias} shows
the relationship between estimator bias, unretrieved prevalence,
and recall.  Bias is greater for low prevalence, small sample size,
and low recall.

\subsection{Exact confidence intervals}
\label{sect:exact-ci}

Instead of a point estimate on recall, we can provide an interval
within which recall falls with a given confidence.
Let $C(X)$ be a function that takes a sample statistic $X$ and
produces an interval $[\lowerbound{\theta}, \upperbound{\theta}]$
on some population parameter $\theta$ (such as recall),
and let $P_{\theta}$ be the probability distribution over 
$X$ for a given value of $\theta$.
Then $C(x)$ (where $x$ is a realization of $X$) 
is a $100 * (1 - \alpha)$ confidence interval if:
\begin{equation}
P_{\theta}\{\theta \in C(X) \} \geq 1 - \alpha \; ,
\label{eqn:ci}
\end{equation}
for all possible values of $\theta$ \cite{lr05:stathyp}.
Note that this equation does not state that there is a $1 - \alpha$
probability that $\theta$ lies within the calculated confidence interval.
In the frequentist understanding, $\theta$ is fixed, not random,
and either lies in the interval, or does not.  
What is random is the sample, $X$; and Equation~\ref{eqn:ci}
states that, whatever the actual value of $\theta$, if we repeatedly
drew random samples and calculated
the confidence interval $C(\cdot)$, then (in the limit)
at least $1 - \alpha$ of these intervals would cover $\theta$.
We refer to $P_{\theta}\{\theta \in C(X) \}$ as the
\defineterm{coverage} of $\theta$ by the confidence interval.

Finite $\lowerbound{\theta}$ and $\upperbound{\theta}$ give a 
two-tailed confidence interval.
Such intervals are often built from a pair of inverted, one-tailed 
hypothesis tests.  Let $x$ be the observed sample statistic, and
$S_l(x ; \alpha/2)$ be the interval 
$[\theta_l, \infty]$ of null hypotheses $\theta_0$ that we accept
at significance level $\alpha/2$
in a lower-tailed hypothesis test, given $x$;
and let $S_u(x ; \alpha/2)$ be the corresponding upper-tailed
interval $[\infty, \theta_u]$.
Then a $1 - \alpha$ confidence interval is defined as 
$S_l(x ; \alpha/2) \cap S_u(x ; \alpha/2) = [\theta_l, \theta_u]$.  
Informally, the
lower end of the confidence interval is the smallest value of
$\theta$ that a lower-tailed hypothesis test fails to reject, 
given the observed
statistic $x$, at level $\alpha/2$, and conversely for the upper end.
We refer to an interval constructed by the inversion of hypothesis
tests that are based on the distribution of the sample statistic 
as an \defineterm{exact} confidence interval.

\begin{figure}
\centering
\includegraphics{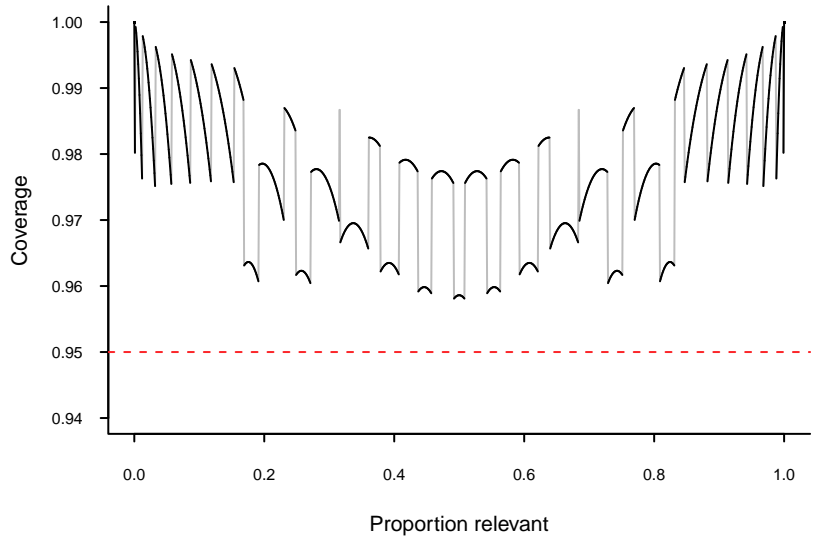}
\caption{Coverage of the exact binomial confidence interval, with
a nominal confidence level of 95\%,
for a sample size of $20$, across different population prevalences.
Discontinuities are shown in gray.}
\label{fig:ex-bin-cvr}
\end{figure}

A confidence interval is a tight one if
${P_{\theta}\{\theta \in C(X) \} = 1 - \alpha}$ for all $\theta$.
Exact intervals are not always tight, and 
exact intervals on discrete sample distributions such as
the binomial generally are not~\cite{neyman35:ams}.
One such exact interval is the Clopper-Pearson exact 
binomial confidence interval on
a binomial proportion, formed from a pair of inverted, one-tailed
binomial significance tests~\cite{cp34:biom}.
Figure~\ref{fig:ex-bin-cvr} shows the coverage of the 
Clopper-Pearson interval at confidence level 95\%,
across different population prevalences, for 
a sample size of $20$\@.  The interval achieves coverage of
at least $1 - \alpha$ for all $\theta$,   
but does not give tight coverage; average coverage
is 97.7\%, and is closer to 99\% for non-mediate prevalences.
(The discontinuities in the coverage of this and subsequently diagrammed
intervals are due to the discrete nature of the sample statistic;
as we vary the population prevalence, particular counts of
positive instances in the sample change from providing intervals 
that cover the population
prevalence to intervals that do not, and vice versa.)

\subsection{Approximate confidence intervals}
\label{sect:approx-ci}

Exact confidence intervals can be difficult to calculate
or analyze, and are conservative for discrete distributions.  
Approximate intervals can be easier to analyze; moreover, while
exact intervals guarantee minimum nominal ($1 - \alpha$) coverage, 
approximate intervals can provide average coverage at the nominal
level, which is preferable for some applications.
(\citeN{lk09:jos} use the term ``coverage
intervals'' for intervals aiming at mean, rather
than guaranteed, nominal coverage.)

The approximate Wald interval on a binomial proportion
calculates the upper and lower bounds as quantiles of a normal distribution
centered upon the sample proportion.  This is equivalent to 
inverting two one-sided normal (Wald) tests, with
standard error estimated from sample prevalence~\cite{ac98:tas}.  If a
size $n$ sample produces sample prevalence $p = r / n$, then sample variance is estimated as:
\begin{equation}
\est{\var}_{p} = \frac{p(1 - p)}{n} \; .
\label{eqn:binom-var-mle}
\end{equation}
(This is a biased estimator; the unbiased
estimate has $(n - 1)$ as its divisor \cite[Chapter 3]{cochran77};
however, Equation~\ref{eqn:binom-var-mle} is
in almost universal use, and is used throughout this article.)
The standard error $s$ is the square root of this variance,
and the confidence
interval is $p \pm s \cdot z_{\alpha/2}$, where $z_{\alpha/2}$ is the 
$\alpha/2$ quantile of the standard normal cumulative distribution
function (for instance, $z_{0.05/2} = 1.96$).  

\begin{figure}
\centering
\includegraphics{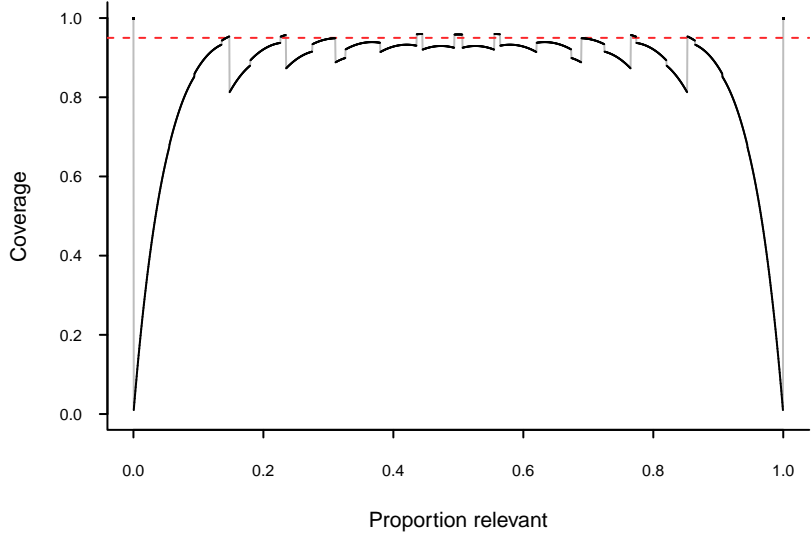}
\caption{Coverage of the Wald binomial confidence interval, 
with a nominal confidence level of 95\%, for a sample size of 
$20$, across different population prevalences. 
Discontinuities are shown in gray.}
\label{fig:bin-wald-ci}
\end{figure}

The Wald interval provides poor coverage, as shown
in Figure~\ref{fig:bin-wald-ci}, where mean coverage 
is 85.1\%, and coverage drops close to 0 for edge cases.
The exact interval is asymmetric when $p \neq 0.5$,
as populations with more extreme prevalences have
lower sampling error than populations with less;
but the Wald interval is always symmetric.  For example, 
the Wald interval for a sample prevalence of $0$
is $[0, 0]$, even though 
a zero prevalence sample can be produced from a population 
with non-zero prevalence.

\begin{figure}
\centering
\includegraphics{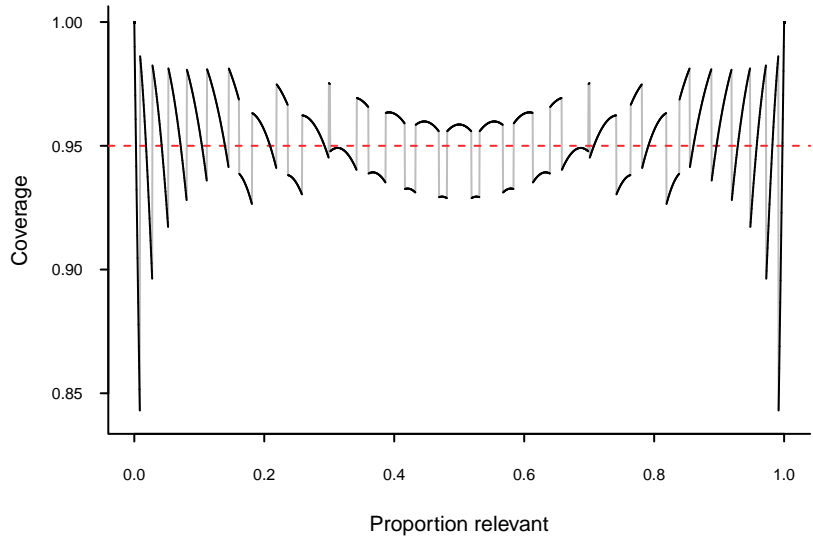}
\caption{Coverage of the Wilson binomial confidence interval, 
with a nominal confidence level of 95\%, for a sample size of 
$20$, across different population prevalences. 
Discontinuities are shown in gray.}
\label{fig:bin-wilson-ci}
\end{figure}

The Wilson (or score) interval on the binomial also
inverts normal hypothesis tests, but uses the prevalences at the
candidate lower and upper bounds to estimate standard
errors.  Thus, the lower half
of the interval is shorter than the upper for $p < 0.5$,
and vice versa for $p > 0.5$, as with the exact interval.
The precise formula is:
\begin{equation}
\frac{p + (z_{\alpha/2}^2 / 2n) \pm z_{\alpha/2}
   \sqrt{[p(1 - p) + z_{\alpha/2}^2/4n]/n}}{1 + z_{\alpha/2}^2/n} 
\label{eqn:wilson-ci}
\end{equation}
\cite{ac98:tas}.
The coverage of the Wilson interval for our example scenario is shown in 
Figure~\ref{fig:bin-wilson-ci}.  Mean coverage here is 95.3\%.

\citeN{ac98:tas} show that a 95\% Wilson confidence interval
is closely approximated by adding two to the count of positive
and of negative items observed in the sample, to form an adjusted
sample size $\tilde{n} = n + 4$ and
sample prevalence $\tilde{p} = (r + 2) / \tilde{n}$, and using the adjusted
$\tilde{p}$ and $\tilde{n}$ in the Wald interval,
to produce an interval of:
\begin{equation}
\tilde{p} \pm z_{0.025} \sqrt{\tilde{p} (1 - \tilde{p}) / \tilde{n}} \; .
\end{equation}
More generally, $z_{\alpha/2}^2 / 2$ positive and $z_{\alpha/2} ^2 / 2$ 
negative observations are added, where $z_{\alpha/2}$ is the 
appropriate quantile
of the standard cumulative normal distribution.

\subsection{Prior and posterior distributions}
\label{sect:bayes}

An alternative approach to confidence intervals is 
offered by the Bayesian viewpoint.
Here, the true parameter is treated as a random
variable, due to the incompleteness of our subjective 
knowledge of its value.  
We posit a distribution over the parameter, 
which describes our subjective belief about
the parameter's likely values, prior to observing the evidence
of a sample.  The prior distribution is updated from
the observed evidence to form a posterior distribution, expressing
our belief given the evidence~\cite{gcsr04:bda}.  
A $1 - \alpha$ confidence interval can be derived
by reading off the $\alpha / 2$ and $1 - (\alpha / 2)$
quantiles of the cumulative posterior distribution.
(Such intervals are known as ``credible intervals'' in Bayesian
statistics~\cite{bolstad07:byst}.)

Let $\theta$ be the parameter of interest, and $x$ the
evidence observed (say, a sample statistic).  Given a prior
probability distribution $p(\theta)$ on $\theta$, we
wish to infer a posterior distribution $p(\theta \given x)$, given 
observation $x$.  Using Bayes' rule, the posterior is re-written as:
\begin{equation}
p(\theta \given x) = \frac{p(x \given \theta)p(\theta)}{p(x)}  
   = \frac{p(x \given \theta)p(\theta)}
     {\int p(x \given \theta)p(\theta) \, \dif \theta}\; .
\end{equation}
That is, the posterior probability of $\theta$ given $x$ is the
prior probability of $\theta$, times the likelihood of $x$ given
$\theta$, divided by the marginal probability of $x$, derived
by integrating over our
prior distribution on $\theta$ \cite{gcsr04:bda}.

For computational and analytic convenience, 
it is common to choose the prior distribution $p(\theta)$
from a family of distributions such that the posterior,
$p(\theta \given x)$, is from the same family, given the
distribution of the likelihood, $p(x \given \theta)$.
We say in this case that the distribution of the 
prior is conjugate to that of the likelihood.
For instance, the beta distribution, $\betadist(\alpha, \beta)$,
with probability distribution function:
\begin{equation}
f(q ; \alpha, \beta) = \frac{\Gamma(\alpha + \beta)}
   {\Gamma(\alpha)\Gamma(\beta)} q^{\alpha - 1} (1 - q)^{\beta - 1}
\end{equation}
is conjugate prior to the binomial.  
($\Gamma(\cdot)$ is the gamma function; for positive integer $n$, 
$\Gamma(n) = (n - 1)!$.)
We use $f(q ; \alpha, \beta)$
to express the prior probability $\Pr(\pi = q)$ that the
true prevalence $\pi$ is $q$.  If the $n$-size sample 
contains $r$ positive instances, and $n - r$ negative instances,
then the prior distribution over $\pi$ of $\betadist(\alpha, \beta)$
is updated to the posterior distribution of $\betadist(\alpha + r,
\beta + n - r)$.

\begin{figure}
\centering
\includegraphics{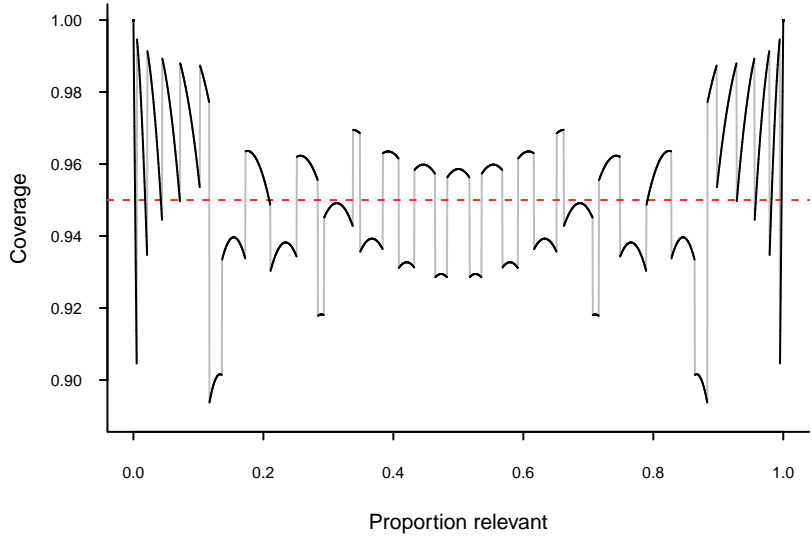}
\caption{Coverage of the binomial confidence interval based upon
a Jeffreys prior, with a nominal confidence level of 95\%, 
for a sample size of $20$, across different population prevalences. 
Discontinuities are shown in gray.
}
\label{fig:bin-jeffreys-ci}
\end{figure}

A particular prior distribution is instantiated from a family
by the choice of hyperparameters, such as $\alpha$ and $\beta$ for
the beta distribution.  In the absence of previous information
about $\theta$, a non-informative prior is generally
chosen, one which is non-committal about the value of $\theta$.
A simple approach for a single parameter of finite
range is to choose hyperparameters that give a uniform prior
distribution over $\theta$, making each value equiprobable.
Setting $\alpha = \beta = 1$ achieves this for the beta prior.
The uniform prior, however, may 
be too strong; for small samples of unbalanced binomial 
populations, it may pull inferences on prevalence $\pi$ too far towards 
$\pi = 0.5$.

A more formally derived prior is the Jeffreys prior, which is
the distribution proportional to (for a single parameter) the
square root of the parameter's expected Fisher information
(a measure of the mean squared error of an estimator,
which is invariant to parameter transformations,
such as converting $\pi$ to a logit scale~\cite{jeff46:prs}).
The Jeffreys prior
for the binomial is $\betadist(0.5, 0.5)$; after observing
$r$ positive and $n - r$ negatives, the resulting
posterior is $\betadist(0.5 + r, 0.5 + n - r)$.

Figure~\ref{fig:bin-jeffreys-ci} shows the coverage of a 95\% 
Jeffreys interval for a sample size of $20$; mean coverage
here is 95.1\%. 
(As the beta is a continuous distribution in the range $[0, 1]$,
the lower quantile is always greater than $0$, and the upper
less than $1$;  we adjust the
interval by setting its
lower end to $0$ if $r = 0$, and the upper end to $1$ if $r=n$).
Along with the Wilson (Section~\ref{sect:approx-ci}), 
the Jeffreys interval has been
recommended as giving the best two-sided coverage for the
binomial~\cite{bcd01:statsci,ac98:tas}.  The one-sided Wilson interval
is biased in $p$ (the lower-bound interval under-covers
for low $p$, and over-covers for high $p$); the Jeffreys interval
avoids this bias~\cite{lk09:jos}.

\subsection{Beta-binomial as prior to hypergeometric}
\label{sect:beta-bin-prior}

Sampling without replacement from a finite 
binomial population is most precisely modelled by
the hypergeometric distribution,
the conjugate prior to which is the beta-binomial.
We express the prior probability, $\Pr(R = s)$, that the 
yield $R$ of a population of size $N$ is $s$, as
$\betabindist(\alpha,\beta;N)$, having the
probability distribution function:
\begin{equation}
g_1(s ; N, \alpha, \beta) = {N \choose s} \frac
   {B(s + \alpha, N - s + \beta)}
   {B(\alpha, \beta)} \; ,
\label{eqn:bb-prior}
\end{equation}
where $B(x, y)$ is the beta function, $\Gamma(x)\Gamma(y)/\Gamma(x + y)$.
Having sampled $n$ elements and observed $r$ positives, the posterior
distribution on $R$ is:
\begin{equation}
g_p(s \given r ; N , n, \alpha, \beta) = {N - n \choose s - r} \frac
   {B(s + \alpha, N - s + \beta)}
   {B(\alpha + r, \beta + n - r)} \; , s = r, r+1, \ldots, N - n + r \;.
\label{eqn:bb-post}
\end{equation}
Note that $g_p(s \given r ; N, n, \alpha,\beta) = g_1(s - r; N - n, \alpha + r,
\beta + n - r)$, establishing conjugacy.

As with the beta prior to the binomial, setting 
$\alpha = \beta = 1$ gives a (discrete) uniform 
prior for $R$ over $[0 \cdots N]$;
but again, this may be too strong for small samples from
unbalanced populations.  A weaker prior (which we will refer to
as the \defineterm{half prior}) sets $\alpha = \beta = 0.5$, as 
with the Jeffreys prior to the binomial, though without the
same theoretical justification: since the discrete
distribution of the yield $s$ is not differentiable,
the Fisher information of the prior cannot be calculated~\cite{bbs08:tr}.

\makeatletter
\let\hypparam\@undefined
\let\prior\@undefined
\let\posterior\@undefined
\makeatother

An alternative non-informative prior 
is the \defineterm{most conservative prior},
which maximizes the expected information gain
from the observed data, as measured by the Kullback-Leibler
divergence between prior and posterior distributions~\cite{dc84:cstm}.
\citeN{dp93:cstm} show that the expected information gain
for the beta-binomial prior  to the hypergeometric is: 
\begin{equation}
\begin{gathered}
\bar{D}_{\mathrm{BB}}(\alpha, \beta; N, n) = \frac{\Gamma(\alpha + \beta)}
   {\Gamma(\alpha)\Gamma(\beta)\Gamma(\alpha + \beta + N))} \\
   \times \sum_{x=0}^n \sum_{r=x}^{N-n+x} {n \choose x}{N -n \choose r - x}
   \Gamma(\alpha + r) \Gamma(\beta + N - r) \\
   \times \ln \left[ \frac{{N - n \choose r - x} \Gamma(\alpha) \Gamma(\beta)
        \Gamma(\alpha + \beta + n)}
        {{N \choose r} \Gamma(\alpha + x) \Gamma(\beta + n - x)
             \Gamma(\alpha + \beta)} \right] \; 
\end{gathered}
\label{eqn:mean-info-gain}
\end{equation}
(correcting 
Equation 3.1.3 of \citeN{dp93:cstm}, which has ${n \choose x}$
instead of ${N \choose r}$ in the final line). 
\citeN{dp93:cstm} also demonstrate that
Equation~\ref{eqn:mean-info-gain} is concave and symmetric in $\alpha$
and $\beta$, so the maximum occurs where $\alpha = \beta$.  Finding
the value $\alpha = \beta$ which maximizes 
Equation~\ref{eqn:mean-info-gain} provides the most 
conservative prior~\cite{dp93:cstm}.
We solve Equation~\ref{eqn:mean-info-gain} as an optimization problem;
solutions fall in the range $[0.1, 1.0]$.
The double sum in Equation~\ref{eqn:mean-info-gain} makes it
$O(Nn)$ to calculate, while for large $N$ and mediate $n/N$, 
the solution in $\alpha = \beta$ appears to converge; we
therefore limit $N$ to $1000$ and $n$ to $1000 - \min(N - n, 200)$.

The most conservative prior has anomalous behavior for tiny
sample sizes.  For $n = 1$, the prior hyperparameters collapse 
to $0$, and the posterior distribution to $p(s) = 1$, where $s = 0$
if the sampled element is negative, and $s = N$ if positive.  For
$n > 1$ but still small, the prior likewise tends to be too weak.
The potential for this to lead to inaccurate interval
coverage is observed empirically in Section~\ref{sect:eval}.

\subsection{Monte Carlo estimation of intervals}
\label{sect:mc-int}

Bayesian confidence intervals are taken from
the quantiles of the posterior cumulative distribution
of the parameter being bounded.  The posterior on recall
is a function of beta or beta-binomial posteriors
(Sections~\ref{sect:jeffreys} and \ref{sect:hypergeom})
on segment yield; deriving a closed-form expression for
the combined distribution is not straightforward.
One could instead calculate the probabilities
of all $(N_0 + 1) \cdot (N_1 + 1)$ yield combinations, 
and the recall for each; sort by recall; and then accumulate
probabilities from each end until the desired quantiles are
achieved.  This is $O(N^2)$ in both time
and memory.  

Instead of exhaustive computation, we estimate
the interval by Monte Carlo 
simulation~\cite{buck84:biom,cs99:jcgs}.
The beta or beta-binomial posteriors on
retrieved and unretrieved yield are
independent, and can be efficiently sampled from~\cite{cheng78:cacm}.
We draw $s$ such samples in pairs,
one from the posterior of each segment, and calculate the recall for
each sample.
These $s$ recall values are sorted, and the $\alpha/2$ and
$1 - \alpha / 2$ quantiles taken as estimates of the
like quantiles of the distribution, and hence 
of the interval.  
For the methods
described in
Sections~\ref{sect:jeffreys} and \ref{sect:hypergeom}, 
we employ $40{,}000$ draws; this takes around $50$ milliseconds on 
a current single processor.

\subsection{Propagation of error}
\label{sect:prop-err}

The Wald and similar approximate intervals are calculated
from a normal distribution centered on the point estimate;
the spread of the distribution depends on the sampling
error of the statistic.
If the sample statistic is a function
of other statistics, the variance of the 
approximation is likewise a function of the component variances.
If a random variable $X$ is a function of other random
variables $A$ and $B$, written:
\begin{equation}
X = f(A,B)  \; , 
\label{eqn:fab}
\end{equation}
then, by the theory of \defineterm{propagation of error}:
\newcommand{\partdifffrac}[2]{\ensuremath{\frac{\partial #1}{\partial #2}}}
\newcommand{\partdiffvar}[2]{\ensuremath{\left( \partdifffrac{#1}{#2} \sigma_{#2} \right)^2}}
\begin{equation}
\var(X) = \partdiffvar{f}{A} + \partdiffvar{f}{B} 
  + 2 \left| \partdifffrac{f}{A} \partdifffrac{f}{B} \right| \cov_{AB} \; ,
\label{eqn:prop-uncert}
\end{equation}
where $\sigma_Y = \sqrt{\var(Y)}$ is the standard deviation of
$Y$, $\cov_{YZ}$ is the covariance between $Y$ and $Z$,
and $\left| Y \right|$ is the absolute value of $Y$
\cite{taylor97:erranl}.  

The recall estimate in Equation~\ref{eqn:est-rec-alt} is 
a function of the variables $\est{R}_1$ and $\est{R}_0$.
Since $\est{R}_0$ and $\est{R}_1$ are sampled from
distinct segments, they are independent (unlike
$\est{R}_1$ and $\est{R}_*$ in Equation~\ref{eqn:est-rec}), so the
covariance term in Equation~\ref{eqn:prop-uncert} is $0$.  
It can be shown that:
\begin{equation}
\partdifffrac{\est{\recall}}{\est{R}_0} = - \frac{1}{\est{R}_1 \left(
      1 + \est{R}_0/\est{R}_1 \right) ^ 2} \; ,
\end{equation}
and that:
\begin{equation}
\partdifffrac{\est{\recall}}{\est{R}_1} = \frac{\est{R}_0}{\left(
      \est{R}_1 + \est{R}_0 \right) ^2} \; .
\end{equation}
Applying Equation~\ref{eqn:prop-uncert}, we find 
that:
\begin{equation}
\est{\var}(\est{\recall}) = \frac{\est{\var}(\est{R}_1) \est{R}_0^2 
  + \est{\var}(\est{R}_0) \est{R}_1^2}
         {(\est{R}_1 + \est{R}_0)^4}  \; ,
\label{eqn:acc-var-est-rec}
\end{equation}
where:
\begin{equation}
\est{\var}(\est{R}_1) = N_1^2 \cdot \est{\var}_{p_1} \cdot
    \left(1 - \frac{n_1}{N_1}\right) \;,
\label{eqn:var-est-seg}
\end{equation}
and similarly for $\est{\var}(\est{R}_0)$;
note the finite-population adjustment in the last term.
A maximum likelihood estimate (MLE) of $\est{\var}_{p_1}$ may be 
taken from Equation~\ref{eqn:binom-var-mle}.
For stratified sampling, the 
segment estimate variance is the sum of stratum estimate variances
(Equation~\ref{eqn:var-est-seg} with stratum values).
\citeN{legal08:trec} propagate error
from Equation~\ref{eqn:est-rec}, instead of
Equation~\ref{eqn:est-rec-alt}, while ignoring the covariance
term in Equation~\ref{eqn:prop-uncert}.
This naive estimate overstates recall variance by
including $\est{\var}(\est{R}_1)$ twice.

\section{Recall confidence intervals}
\label{sect:rec-ci}

\begin{table}
\tbl{Recall confidence intervals.  \label{tbl:rec-ci}}{
\newcommand{\nl}{\\ \addlinespace[0.5em]}
\begin{tabular}{lc>{\raggedright\arraybackslash}p{0.36\textwidth}>{\raggedright\arraybackslash}p{0.24\textwidth}}
\toprule
Name & Section & Summary & Coverage \\
\midrule
Naive binomial & \ref{sect:naive-binom} 
               & Treat sensitivity as binomial proportion 
                   sampled from true positives. 
               & Very inaccurate in most conditions. \nl
Normal MLE     & \ref{sect:normal-mle}
               & Normal approximation with maximum-likelihood
                   variance estimate, aggregated by propagation
                   of error. 
               & Unstable; biased for very low prevalences
                   or heavy samples. \nl
Normal Laplace & \ref{sect:normal-adj} 
               & Add one to positive and negative sample counts in each
                   segment, then as per Normal MLE. 
               & Slightly less unstable and biased than Normal MLE. \nl
Normal Agresti-Coull  & \ref{sect:normal-adj} 
               & Add two to sample counts, then as per Normal MLE.
               & As unstable and biased as Normal MLE. \nl
Koopman        & \ref{sect:koopman}
               & Monotonic decreasing transform of interval on ratio 
                   between binomial variables.
               & Biased to overcoverage for heavy samples; otherwise
                   accurate.\nl
Beta Jeffreys  & \ref{sect:jeffreys}
               & Infer beta posteriors from Jeffreys priors on segments;
                   Monte Carlo generation of recall distribution; take
                   quantiles of generated distribution. 
               & Biased to undercoverage for heavy samples; otherwise
                   accurate.\nl
Beta-binomial Uniform & \ref{sect:hypergeom}
               & Infer beta-binomial posteriors from uniform
                    priors on segments; then as per Beta Jeffreys. 
               & Slightly unstable for very low prevalences, marginally
                    understates recall; otherwise accurate.\nl
Beta-binomial MCP & \ref{sect:hypergeom}
               & Infer beta-binomial posteriors from most conservative
                    priors on segments; then as per Beta Jeffreys.
               & Slightly unstable, overstates recall in some conditions;
                    otherwise accurate.\nl
Beta-binomial Half & \ref{sect:hypergeom}
               & Infer beta-binomial posteriors from priors 
                    with hyperparameters $\alpha = \beta = 0.5$ 
                    on segments; then as per Beta Jeffreys. 
               & Accurate in all conditions examined.\nl
\bottomrule
\end{tabular}
}
\end{table}

Using the techniques laid out in Section~\ref{sect:prelim},
this section proposes nine recall intervals, summarized in
Table~\ref{tbl:rec-ci} (coverage results n the fourth column look forward to
Section~\ref{sect:eval}).   The first four methods use normal
approximations; the fifth, an analytic interval
on a ratio between binomials; and the last four,
Bayesian segment posteriors (the first from the beta prior to the
binomial, and the next three from beta-binomial priors to the
hypergeometric) and Monte Carlo simulation.

\subsection{Sensitivity as a binomial proportion}
\label{sect:naive-binom}

The sensitivity of a diagnostic procedure, like the
recall of a classifier, is the proportion of positive
instance correctly identified.
\citeN{ssm91:jclep}
treat sensitivity as a binomial proportion on the population of
true positives, and infer a Wald interval
on this proportion (Section~\ref{sect:approx-ci}),
to give a $1 - \alpha$ recall (or sensitivity) confidence interval of:
\begin{equation}
\est{\recall} \pm z_{\alpha/2} \cdot 
    \sqrt{\est{\recall} (1 - \est{\recall}) / n} \; .
\label{eqn:ci-binom}
\end{equation}
We refer to this as the \defineterm{naive binomial}
interval.

The naive binomial assumes that positive instances are
equally sampled, which, though valid in the 
epidemiological randomized control trials considered 
by \citeN{ssm91:jclep}, 
does not generally hold for retrieval recall estimation,
since the retrieved and unretrieved segments may be 
sampled separately, and at different rates.  The same
inequality in sampling rate means the interval extends
poorly to stratified sampling.

\subsection{Normal with maximum-likelihood variance}
\label{sect:normal-mle}

The sampling variance of recall can be derived
from segment yield variance by error propagation
(Equation~\ref{eqn:acc-var-est-rec}), and
standard error is the square root of variance:
\begin{equation*}
\est{s}_R = \sqrt{\est{\var}(\est{\recall})}
\end{equation*}
If we assume that $\est{\recall}$ is normally
distributed, and has equal variance along its range, 
then a $1 - \alpha$ confidence interval is:
\begin{equation}
\est{\recall} \pm z_{\alpha/2} \cdot \est{s}_{\mathrm{R}} \; .
\label{eqn:ci-normal-mle}
\end{equation}
Note that the interval is centered on the
recall point estimate.  A direct estimate
of $\var(\est{\recall})$ uses $\est{\var}_p = p (1 - p) / n$
(Equation~\ref{eqn:binom-var-mle}) in Equation~\ref{eqn:var-est-seg}.
We refer to the
approximate normal interval with this estimates
as the Normal MLE method.  Stratified sampling is 
handled by summing stratum variances, 
as described in Section~\ref{sect:prop-err}.

The Normal-MLE method has two problems.  First,
the sampling distribution of recall may diverge from
the normal (Section~\ref{sect:rec-dist}); for instance, 
recall is bounded by the range $[0, 1]$, whereas
the normal distribution is unbounded.  And second, the
MLE of $\var(\est{R})$ is biased low for 
extreme sample prevalence.
In particular, if $p_0 = 0$, then $\est{\var}(\est{R}_0)$ is $0$ and the
recall interval is $1 \pm 0$, but perfect recall is
certainly not assured.

\subsection{Laplace and Agresti-Coull adjustments to the normal}
\label{sect:normal-adj}

\citeN{ac98:tas} adjust the 
Wald confidence interval on the binomial proportion 
by adding $2$ to the count of
positive and negative sample instances
(Section~\ref{sect:approx-ci}).
A similar idea can be applied to the interval on recall,
by adding $1$ or $2$ to the positive and negative sample
counts in the retrieved and unretrieved segments. 
We denote the addition of $2$
as the Agresti-Coull adjustment, and of $1$ as the
Laplace adjustment (from Laplace's prior to the 
binomial) \cite{ac00:tas,greenland01:tas}.
In stratified sampling, the adjustment would be applied
to each stratum.

If $c$ is the amount of the adjustment, then the adjusted
proportion relevant for each segment or stratum, 
$\tilde{p} = (r + c) / (n + 2c)$,
is used in Equation~\ref{eqn:binom-var-mle} to derive
an adjusted expression for $\est{\var}_p$ in 
Equation~\ref{eqn:var-est-seg}~\cite{ac98:tas}:
\begin{equation}
\widetilde{\var}_{p} = \frac{\tilde{p}(1 - \tilde{p})}{n + 2c} \; .
\label{eqn:adj-binom-var-mle}
\end{equation}
The adjusted variance expression is then propagated through 
Equation~\ref{eqn:acc-var-est-rec}, with adjusted estimators
of retrieved and unretrieved yield:
\begin{equation}
\tilde{R}_1 = N_1 \tilde{p}_1 \quad ; \quad \tilde{R}_0 = N_0 \tilde{p}_0  \; ,
\end{equation}
to calculate an adjusted variance ($\widetilde{\var}(\est{\recall})$)
and standard error 
($\tilde{s}_{\mathrm{R}} = \widetilde{\var}(\est{\recall})^{1/2}$) 
for recall.
The midpoint for the interval is also calculated with the
adjusted counts:
\begin{equation}
\widetilde{\recall} = \frac{\tilde{R}_1}{\tilde{R}_1 + \tilde{R}_0} \; .
\end{equation}
And finally, the adjusted interval is estimated with these adjusted
values:
\begin{equation}
\widetilde{\recall} \pm z_{\alpha/2} \cdot \tilde{s}_{\mathrm{R}} \; .
\label{eqn:ci-normal-adj}
\end{equation}
These adjustments make
the interval more robust to extreme sample prevalences.
We force the top end of the confidence interval to $1$ if there
are no positives in the unretrieved sample, and the bottom to $0$
if none in the retrieved.

\subsection{Ratio between binomial proportions: Koopman}
\label{sect:koopman}

\citeN{koopman84:biom} derives
an approximate confidence interval on the ratio of two 
independent binomial
proportions, which he refers to as the chi-square method.
(The equations are too lengthy to reproduce here; see
Section 2.2 of \citeN{koopman84:biom} for details.)
The formula for recall in Equation~\ref{eqn:est-rec-alt} 
is a monotonically decreasing function
of the ratio between the independent binomial proportions
$\est{R}_0$ and $\est{R}_1$.  Therefore, a $1 - \alpha$
confidence interval on the latter ratio will also be
a $1 - \alpha$ confidence interval on recall,
with ends reversed~\cite{smithson02:ci}.
We
refer to this as the Koopman interval.
We set the lower bound to $0$ if there are no relevant documents
in the retrieved sample, and the upper bound to $1$ if there are
no relevant documents in the unretrieved sample.
The method does
not have a straightforward extension to stratified sampling,
since the compound samples from retrieved and unretrieved 
segments cease being simple binomial random samples.

\subsection{Ratio between beta posteriors with Jeffreys priors}
\label{sect:jeffreys}

A beta posterior based on a Jeffreys prior provides a
binomial confidence interval with good coverage
(Section~\ref{sect:bayes}).  The method can be adapted to derive
an interval on recall.  Beta posteriors are inferred
on the prevalences of the
retrieved and unretrieved segments, $\pi_1$ and $\pi_0$.
A pair of independent observations, 
$\pi_1^*$ and $\pi_0^*$, 
is drawn from these posteriors, 
and the yields for the corresponding segments are calculated.  For
instance, for the retrieved segment:
\begin{gather*}
\pi_1^* \leftarrow \betadist(0.5 + r_1, 0.5 + n_1 - r_1)  \\
R_1^* = r_1 + \pi_1^* \cdot (N_1 - n_1) \; .
\end{gather*}
(Note the adjustment for sample size in $N_1 - n_1$, and for 
sample yield in $r_1$; we have seen a portion of the population
in the sample, and do not need to estimate the yield of that
portion.)  A recall value is then
observed from these segment yield estimates.  The process
is repeated multiple times, and a $1 - \alpha$ interval is taken from
the $\alpha / 2$ and $1 - \alpha/2$ quantiles of the resulting 
Monte Carlo distribution (Section~\ref{sect:mc-int}).
The lower bound is forced to $0$ if there are no relevant
documents in the retrieved sample, and the upper bound to $1$
if there are none in the unretrieved sample.
We refer to this as
the Beta Jeffreys recall interval.  Stratified sampling is handled
by separately deriving posteriors for each stratum, and drawing
samples from each such posterior.

\subsection{Ratio between beta-binomial posteriors with uniform,
most conservative, or half priors}
\label{sect:hypergeom}

Both Koopman and Beta Jeffreys intervals are based on
the binomial distribution.  In practice, though, the
document population is finite, and the hypergeometric
distribution is the more accurate model, the conjugate
prior to which is the beta-binomial (Section~\ref{sect:beta-bin-prior}).
For our final approximate recall interval, beta-binomial
posteriors are inferred for the yields of the retrieved and unretrieved
segment, $R_1$ and $R_0$, and a Monte Carlo
distribution over recall is generated by drawing paired observations
from these posteriors.  For instance,
\begin{equation*}
R_1^* \leftarrow \betabindist(\alpha + r_1, \beta + n_1 - r_1)  
\end{equation*}
The $\alpha / 2$ and $1 - \alpha / 2$ quantiles of the
generated distribution provide a $1 - \alpha$ confidence
interval on recall.

Section~\ref{sect:beta-bin-prior} considers three choices for 
the $\alpha$ and $\beta$ hyperparameters to the beta-binomial
prior. 
The first sets $\alpha = \beta = 1$, giving a uniform
prior (Beta-binomial Uniform).
The second selects the information-theoretic most conservative prior
for the population and sample size (Beta-binomial MCP).
And the third mimics the
Jeffreys prior to the binomial, by setting $\alpha = \beta = 0.5$
(Beta-binomial Half).
Stratified sampling is handled by drawing from posteriors 
on each stratum; the greater number and smaller size of
strata emphasizes the importance of careful prior choice.
The lower bound is forced to $0$ if there are no relevant
documents in the retrieved sample, and the upper bound to $1$
if there are none in the unretrieved sample.

\section{Evaluation}
\label{sect:eval}

This section provides the experimental evaluation of
the intervals described in Section~\ref{sect:rec-ci}.
Intervals are tested for their coverage of
three representative retrieval or classification estimation scenarios
(Section~\ref{sect:eval-meth});  
the beta-binomial posterior with half prior
is found to give the most unbiased, 
balanced, and stable coverage of the methods
considered (Section~\ref{sect:eval-cover}).  We offer 
advice on sampling design, the allocation of assessments
to retrieved and unretrieved segments, and the interval
widths that can be expected (Section~\ref{sect:design}).
Finally, we calculate half-prior beta-binomial intervals for
participants in the Interactive Task of the Legal
Track in TREC 2008 and TREC 2009, and compare them
with the officially reported, normal approximation intervals
(Section~\ref{sect:trec-int}).

\subsection{Evaluation methodology}
\label{sect:eval-meth}

We propose four criteria for evaluating the coverage of an
approximate (two-tailed) confidence interval:
\begin{description}
\item[Unbiasedness] Is mean coverage at the nominal (that is, $1 - \alpha$) 
  level?
\item[Consistency] Is variability in coverage small?
\item[Balance] Do non-covered parameter values fall below and above
  the interval with equal frequency?
\item[Narrowness] Subject to the other criteria, is the interval
width narrow?
\end{description}
An interval satisfying the above criteria will be described
as \defineterm{accurate}.\footnote{Terminology is applied
  differently to exact confidence intervals:  an
  ``unbiased'' interval is as likely to cover the true parameter value
  as any other value~\cite{guen71:tas}, and a ``universally most
  accurate'' interval is one constructed from universally most
powerful hypothesis tests~\cite{lr05:stathyp}.}

With no constraint on width,
a method that gave an interval of $[0, 1]$ 95\%
of the time, of $(0, 0)$ 2.5\% of the time, and of $(1, 1)$
2.5\% of the time, would achieve perfect accuracy, satisfying
the criteria of unbiasedness, consistency, and 
balance.
Narrow intervals, however, can easily be achieved
through inadequate coverage; and unbalanced intervals on
asymmetric distributions can
be narrower by shifting coverage from the longer to the shorter
tail.  Therefore, the width criterion is subject to the other criteria.  

\begin{table}
\tbl{Scenario variables. \label{tbl:scen-vars}}{
\begin{tabular}{lll}
\toprule
Variable & Meaning \\
\midrule
$N_*$            & Size of population \\
$\pi_* = R_*/N_*$    & Proportion relevant in population \\[1em]
$\recall = R_1/R_*$      & Recall of retrieval \\
$\precision = R_1/N_1$ & Precision of retrieval \\[1em]
$n_1$          & Size of sample from retrieved segment \\
$n_0$          & Size of sample from unretrieved segment \\
\bottomrule
\end{tabular}
}
\end{table}

The measured accuracy of coverage depends upon the distribution of population 
parameters.  
In Section~\ref{sect:approx-ci}, we evaluated
binomial intervals using a uniform distribution over prevalence, and
a fixed sample size.  However,
the one recall value can result from different combinations
of population and retrieved segment size and prevalence, and
different sample sizes will provoke different interval behaviors.
Exhaustively searching the full parameter space is infeasible
(and, for population size, formally impossible), and would 
obscure behavior in environments of particular interest.
Instead, we evaluate recall intervals under different retrieval 
or classification \defineterm{scenarios},
each of which provides distributions over population size, prevalence,
and retrieval precision and recall, as well as sample sizes for
the retrieved and unretrieved segments.

Table~\ref{tbl:scen-vars} lists the scenario variables.
A sample from these variables creates a scenario \defineterm{realization}.
We first determine the characteristics of the population or corpus: the
number of documents it contains ($N_*$), and the proportion
of these documents which are relevance ($p_*$).  Then, we determine
the characteristics of the production or retrieval: the proportion
of relevant documents retrieved ($\recall$), and the proportion
of retrieved documents that are relevant ($\precision$).  Finally,
we determine the characteristics of the sample: the number of
relevant documents sampled from the retrieved ($n_1$) and
unretrieved ($n_0$) segments.  All of the values 
in Table~\ref{tbl:notation} can be calculated from these
scenario variables, save the sample outcomes $r_0$ and $r_1$ themselves.
For instance, $N_1$, the retrieval segment size, is 
$\lfloor N \cdot \recall / \precision \rceil$ ($\lfloor \cdot \rceil$
is the rounding operator), while $R_1$, the number of
relevant documents in the retrieved segment, is 
$\lfloor N \cdot \precision \rceil$.
Coverage of each realization is estimated by drawing
repeated samples from the population (simulated by observing
a value for $r$ from a $\hypergeomdist(N, R, n)$ distribution),
and counting the proportion of the interval estimates generated from
these samples that cover the true recall of the realization.

We define three scenarios using these variables, described below:
the neutral, legal, and small scenarios.  The neutral scenario
(Section~\ref{sect:neut-scen}) tests a wide range of conditions, 
avoiding only edge cases of
very low corpus prevalence and sample sizes that are a substantial
proportion of corpus size.  The legal scenario (Section~\ref{sect:lgl-scen})
then tests the large corpus, low-prevalence circumstances encountered in the
TREC Legal Track (and many other retrieval tasks), while the
small scenario (Section~\ref{sect:sml-scen}) investigates the finite 
population case, where sample size is a large proportion of population size.

These three scenarios cover most conditions that are
likely to be encountered in sampling for recall estimation on a
closed corpus, either in a retrieval or classification context, and
test important edge cases that stress
the performance of interval estimators.  (One might observe larger
corpora and larger samples sizes than in the neutral scenario, combined
with more mediate prevalence than in the legal scenario, but
these are unstressful conditions for any estimator.)  Any particular
estimation task will have narrower parameters than the intentionally
broad ones of these scenarios, and so observe difference distributions
of interval accuracy.  A system evaluator might employ the
evaluation methods presented here to test 
interval methods in his or her expected estimation environment.  Nevertheless,
an interval that performs well on all of the realizations of
a broad scenario will perform well on any particular narrower
subset scenario, while intervals that often perform poorly on
the broader case will perform well on only certain of the narrower
ones.

\subsubsection{Neutral scenario}
\label{sect:neut-scen}

\begin{table}
\tbl{Neutral scenario. \label{tbl:neut-scen}}{
\begin{tabular}{cll>{\raggedright\arraybackslash}p{0.45\textwidth}}
\toprule
Variable & Range & Mean & Distribution \\
\midrule
$N_*$       
          & $1{,}000$ to $4{,}000{,}000$
          & $2{,}000{,}500$ 
          & $\unifdist(1e3, 4e6)$ 
\\
$\pi_*$   
          & $0.02$ to $0.8$ 
          & $0.41$ 
          & $\unifdist(0.02, 0.8)$ 
\\
$\recall$   
          & $0.1$ to $1.0$ 
          & $0.55$
          & $\unifdist(0.1, 1.0)$ 
\\
$\precision$ 
          & $0.1$ to $1.0$ 
          & $0.64$
          & $\unifdist(\min(0.1, 0.95 \pi_*, 1.05 \cdot R_1 / N_*), 1.0)$
\\
$n_1$     
          %& $10$ to $\num{4e3}$ 
          & $10$ to $4{,}000$ 
          & $1{,}935$  %%% over sample of 100,000 realizations, +- 7
          & $\unifdist(10, \min(4000, \lfloor N_1 / 10 \rfloor))$
\\
$n_0$     
          %& $10$ to $\num{4e3}$ 
          & $10$ to $4{,}000$ 
          & $1{,}995$ %%% over sample of 100,000 realizations, +- 7
          & $\unifdist(10, \min(4000, \lfloor N_0 / 10 \rfloor))$
\\
\bottomrule
\end{tabular}
}
\end{table}

The neutral scenario is described in Table~\ref{tbl:neut-scen}; 
it represents a variety of classification and retrieval
tasks, on populations large enough to require sampling.  
The neutral scenario covers a wide range
of population prevalences; very low prevalences are saved
for the legal and small scenarios.  All levels of precision
and recall are considered, apart from those so low as to
represent an extremely flawed classification or retrieval.
Two lower bounds are set on precision, the first 
to ensure that retrieved prevalence is not much worse than unretrieved (the
retrieval does at least as well as random), and the second
to guarantee that not all documents are retrieved.   Sample
sizes have equal distributions between retrieved and unretrieved
segments.

\subsubsection{Legal scenario}
\label{sect:lgl-scen}

\begin{table}
\tbl{Legal scenario. \label{tbl:lgl-scen}}{
\begin{tabular}{cll>{\raggedright\arraybackslash}p{0.35\textwidth}}
\toprule
Variable & Range & Mean & Distribution \\
\midrule
$N_*$      
           %& $\num{5e5}$ to $\num{5e7}$ 
           & $500{,}000$ to $50{,}000{,}000$ 
           %& $\num{1.1e6}$
           & $1{,}075{,}000$
           & $\num{5e5}    \cdot 10  ^ {\unifdist(0, 2)}$    
\\
$\pi_*$    
           & $0.003$ to $0.115$ 
           & $0.031$
           & $\num{2e-3}   \cdot 1.5 ^ {\unifdist(1, 10)}$   
\\
$\recall$ 
           & $0.0025$ to $0.84$ 
           & $0.33$
           & $\num{2.5e-3} \cdot \unifdist(1, 34) ^ {1.65}$ 
\\
$\precision$ 
           & $0.025$ to $0.92$ 
           & $0.48$
           & $\unifdist(\min(0.025, 2 \cdot R_1/N_*), 0.92)$
\\
$n_1$     
           & $20$ to $5{,}120$ 
           & $820$
           & $20 \cdot 2 ^ {\unifdist(0, \max(8, \lfloor \log_2 (N_1 / 20) \rfloor))}$
\\
$n_0$      
           & $100$ to $12{,}800$ 
           & $3{,}170$
           & $100 \cdot 2 ^ {\unifdist(0, \max(7, \lfloor \log_2 (N_0 / 100) \rfloor))}$
\\
\bottomrule
\end{tabular}
}
\end{table}

\begin{figure}
\centering
\includegraphics{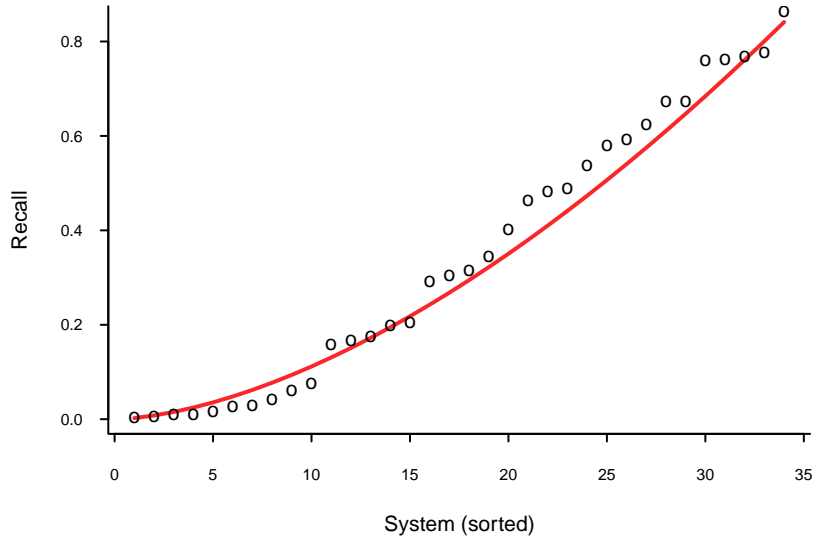}
\caption{Sorted recall of participants in the Interactive Task of the TREC 2008
and TREC 2009 Legal Track (dots), with fitted line of
$\num{2.5e-3} \cdot \unifdist(1, 34) ^ {1.65}$.  Recall scores are sample-based
estimates using adjudicated assessments~\cite{legal08:trec}.
}
\label{fig:rec-fit}
\end{figure}

The legal scenario (Table~\ref{tbl:lgl-scen}) represents 
e-discovery and similar large scale retrieval.
The distributions of scenario variables are fitted to 
parameters observed in the TREC 2008 and TREC 2009 Legal Interactive
task (Section~\ref{sect:trec-int}).
For instance, Figure~\ref{fig:rec-fit} shows the
fit of the recall distribution in Table~\ref{tbl:lgl-scen} to the
recall values from the task.
The scenario corpus size is large, and prevalence low for
a classification (though not for a retrieval) task.  A median
of 2\% of the retrieved segment is sampled, and only 0.5\% of the
unretrieved.  Unretrieved samples
will frequently contain few or no relevant documents, testing
handling of extreme proportions.
The lower bound on precision ensures that no more than
half the corpus is retrieved, and also that very low precision
is not matched with very high recall---the only relationship observed
between the two metrics amongst TREC Legal Interactive participants.

\subsubsection{Small scenario}
\label{sect:sml-scen}

\begin{table}
\tbl{Small scenario.  \label{tbl:small-scen}}{
\begin{tabular}{cll>{\raggedright\arraybackslash}p{0.45\textwidth}}
\toprule
Variable & Range & Mean & Distribution \\
\midrule
$N_*$     
         %& $\num{1e3}$ to $\num{1e4}$ 
         & $1{,}000$ to $10{,}000$ 
         %& $\num{3.9e3}$
         & $5{,}500$
         & $\unifdist(1e3, 1e4)$
\\
$\pi_*$   
         & $0.02$ to $0.22$ 
         & $0.12$
         & $\unifdist(0.02, 0.22)$ 
\\
$\recall$ 
         & $0.1$ to $1.0$ 
         & $0.55$ 
         & $\unifdist(0.1, 1.0)$ 
\\
$\precision$ 
         & $0.025$ to $0.92$ 
         & $0.51$
         & $\unifdist(\min(0.025, 2 \cdot R_1 / N_*), 0.92)$ 
\\
$n_1$     
         & $20\%$ to $50\%$ of retrieved segment 
         & $290$ %% on sample of 100,000; +- 2
         & $N_1 \cdot \unifdist(0.2, 0.5)$
\\
$n_0$     
         & $5\%$ to $30\%$ of unretrieved segment 
         & $815$ %% on sample of 100,000; +- 3.5
         & $N_0 \cdot \unifdist(0.05, 0.3)$
\\
\bottomrule
\end{tabular}
}
\end{table}

The small scenario, described in Table~\ref{tbl:small-scen}, 
represents a retrieval-style task on a small corpus with a
relatively large sample budget.  Sample sizes
are large relative to population, up to 50\% for the retrieved
segment, and up to 30\% for the unretrieved.
Prevalence is allowed to exceed the legal scenario, but is
kept below $0.22$, in line with most retrieval
tasks.  No more than half the population is retrieved.

\subsection{Coverage of the three scenarios}
\label{sect:eval-cover}

We now report the coverage of the nine confidence
interval estimators on the three scenarios described
above, testing the four criteria established at
the start of Section~\ref{sect:eval}: unbiasedness; consistency; 
balance (in Section~\ref{sect:gap}); 
and width (in Section~\ref{sect:width}).

\subsubsection{Neutral scenario}

\begin{figure}
\centering
\includegraphics{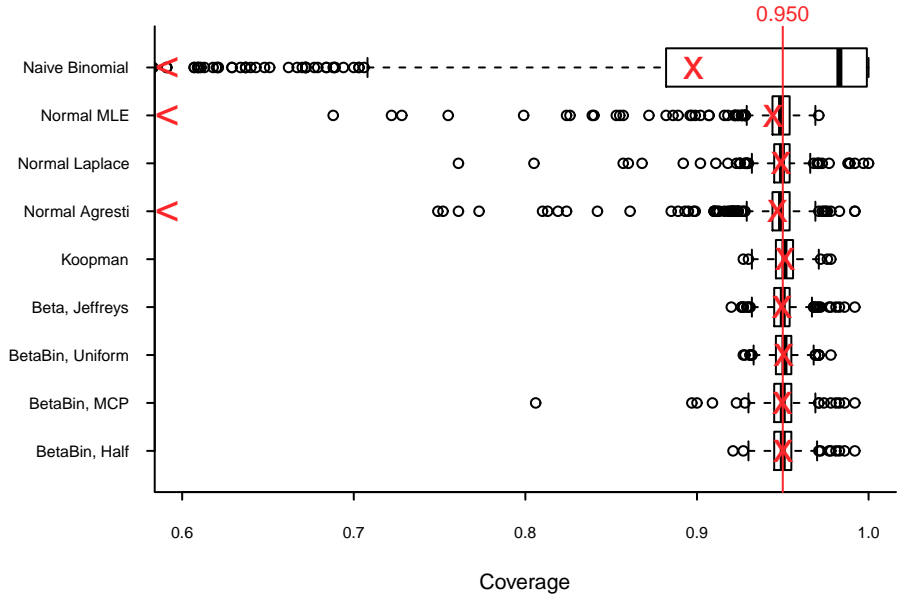}
\caption{Interval coverage for the neutral scenario. 
$1{,}000$ scenario realizations
are instantiated, and $1{,}000$ samples
drawn for each realization; the same realizations and
samples are used for all interval methods.
The thick line in each box shows median coverage, 
the box edges the quartiles.  
Mean coverage is shown as a cross. 
Circles show realizations with coverage
outside the quartiles by more than
$1.5$ times the inter-quartile range.  
Minimum coverage falling below the range of the graph 
is marked on the left margin with ``$<$''.
}
\label{fig:cvr-box-neut}
\end{figure}

Figure~\ref{fig:cvr-box-neut} shows the mean and median 
coverage and coverage consistency of the different
interval methods for the neutral scenario.
The naive binomial is highly inconsistent and biased 
low; its assumptions are violated by different retrieved
and unretrieved sampling rates
(Section~\ref{sect:naive-binom}).  The MLE normal is
biased only slightly low, but is highly unstable, with
frequent significant under-coverage;
the cause is false symmetry and understatement of
variance (Section~\ref{sect:normal-mle}), particularly
for low prevalence unretrieved samples.
The Laplace adjustment corrects the mean bias, 
and avoids the worst cases of instability, but still
has occasions of marked undercoverage,
while the Agresti-Coull adjustment is little
better than the MLE, suggesting the plus-two adjustment
is too large.  The
remaining five methods are nearly unbiased and more
consistent, except that the most-conservative prior to
the beta-binomial shows a tendency
to occasional undercoverage (the prior is too weak for
certain extreme proportions).  The Koopman and
uniform beta-binomial methods give the tightest
behavior.

\subsubsection{Legal scenario}

\begin{figure}
\centering
\includegraphics{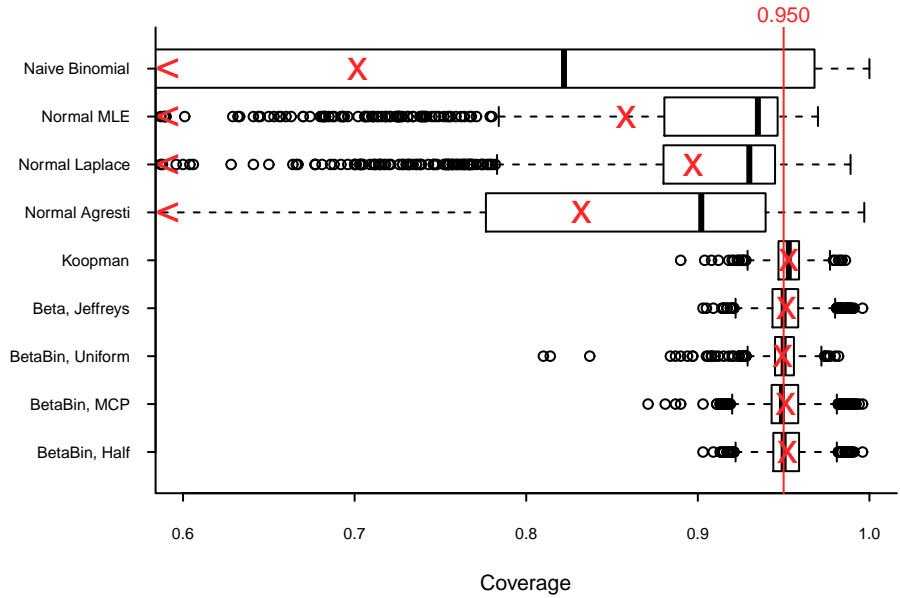}
\caption{Interval coverage for the legal scenario.  
Other details are as for Figure~\ref{fig:cvr-box-sml}.}
\label{fig:cvr-box-lgl}
\end{figure}

The results for the legal scenario are shown
in Figure~\ref{fig:cvr-box-lgl}.  The naive binomial
and normal methods perform much worse than for the
neutral scenario, due to the low prevalence, particularly
in the unretrieved segment.  The Laplace adjustment to the normal
only partially corrects bias, 
while the Agresti-Coull adjustment makes it worse.
Analysis shows that under-coverage is due to over-adjustment:
unretrieved yield is over-estimated, and hence recall understated.
The remaining five methods all give unbiased coverage,  
though this time the uniform prior to the beta-binomial shows
a tendency towards under-coverage, as the uniform prior is too
strong for very low prevalences.

\subsubsection{Small scenario}
\label{sect:eval-sml}

\begin{figure}
\centering
\includegraphics{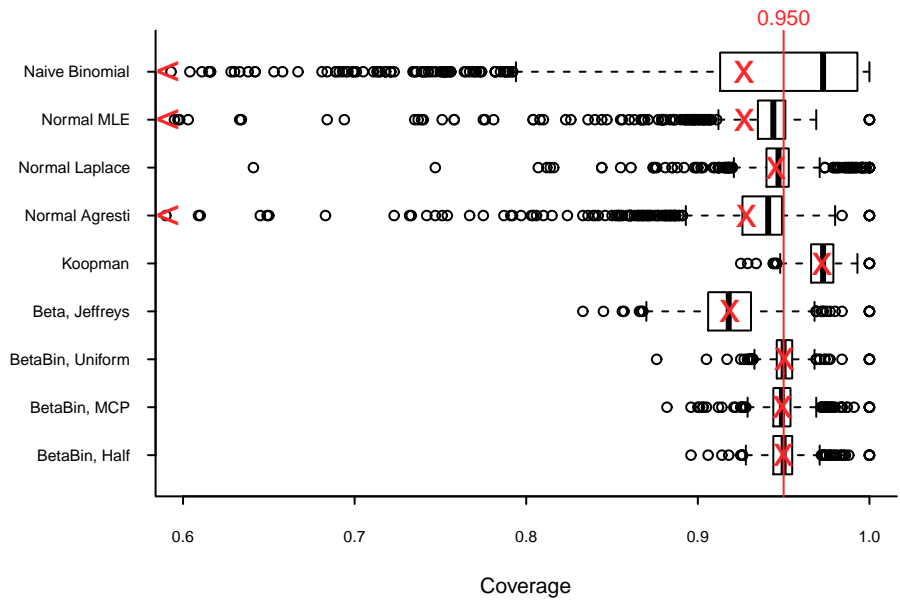}
\caption{Interval coverage for the small scenario.
Other details are as for Figure~\ref{fig:cvr-box-sml}.}
\label{fig:cvr-box-sml}
\end{figure}

The results for the small scenario are given in Figure~\ref{fig:cvr-box-sml}.
The naive binomial and the three normal
methods again show bias and poor consistency, with the Laplace-adjusted
normal being the best of them.  
Unlike the previous scenarios,
however, the Koopman and Jeffreys methods are strongly
biased.  Koopman intervals take no account of the elements
already seen, and are too wide; Jeffreys intervals miss the
negative dependence between sample and unsampled prevalence,
and are too narrow.
The beta-binomial posterior methods are largely accurate, with
the half prior being (marginally) the most consistent of the three.

\subsubsection{Upper and lower gaps}
\label{sect:gap}

\begin{figure}
\centering
\includegraphics{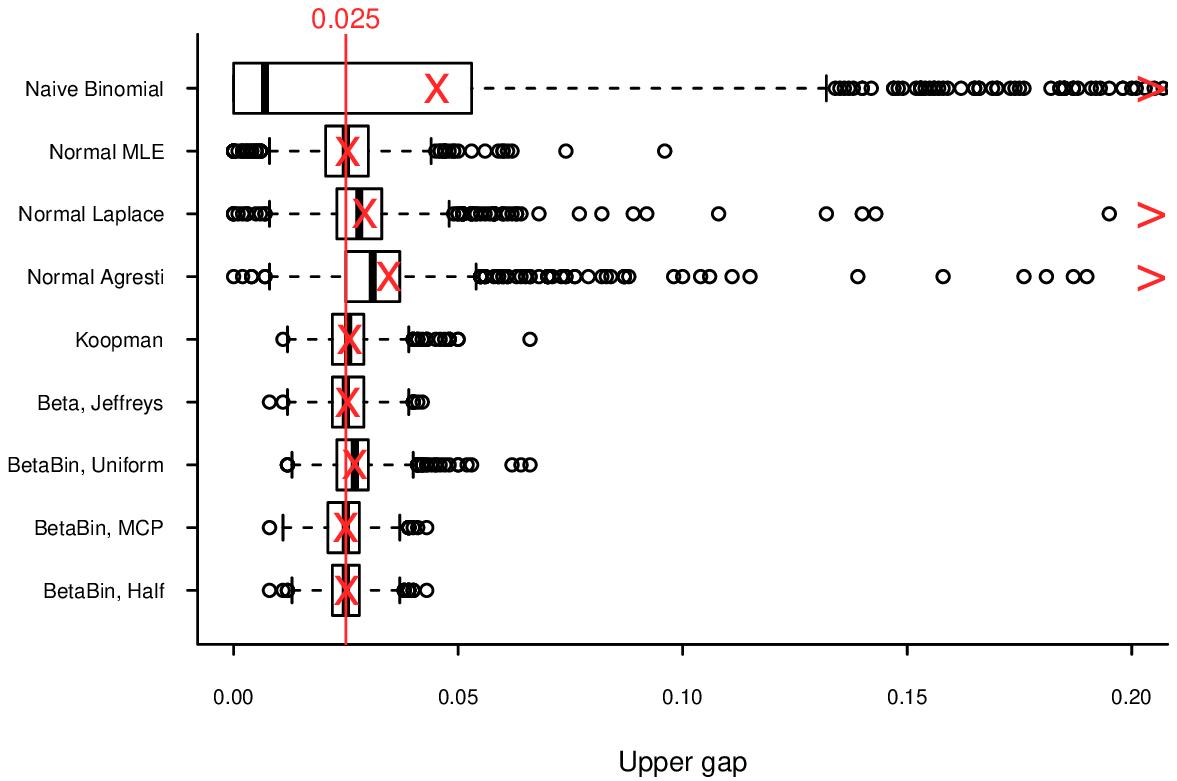} \\
\includegraphics{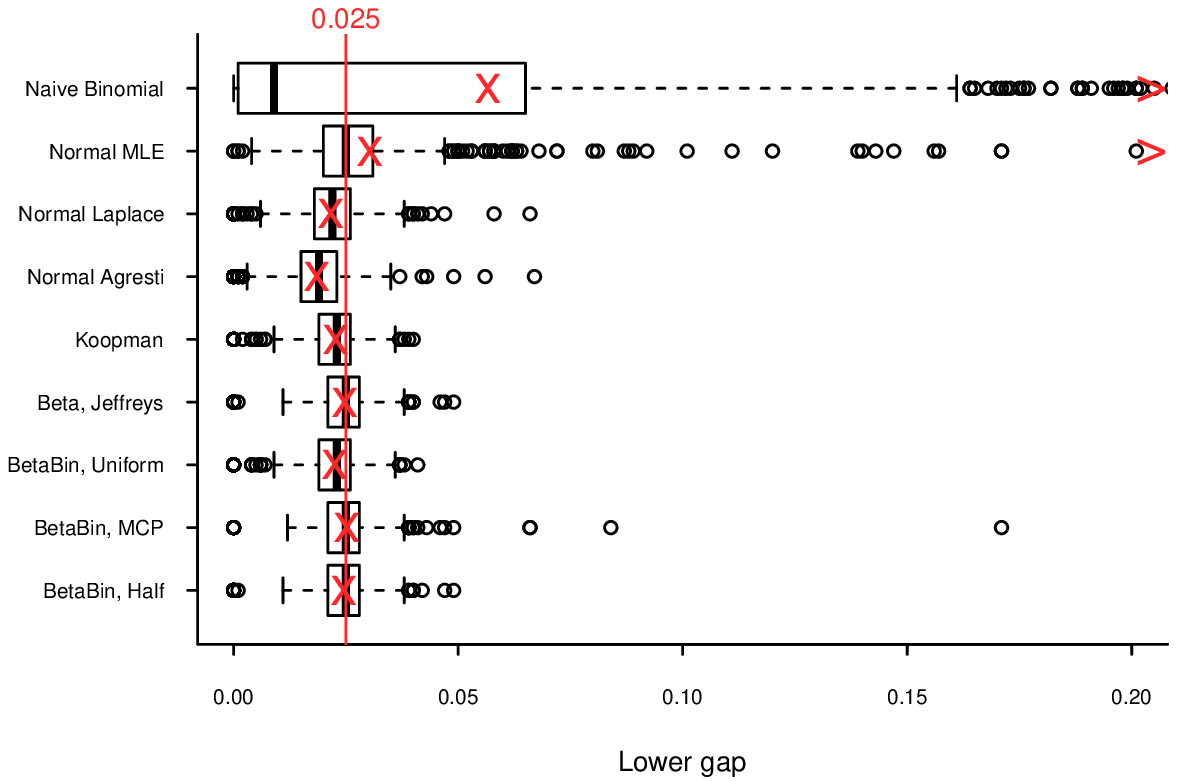} \\
\caption{Upper and lower gaps for different CI methods on the
neutral scenario.  
Each underlying data point represents a scenario realization,
and states the proportion of the samples drawn from that realization
for which recall falls above (or below) the interval estimated from the
sample.  Realizations and resamples are as for Figure~\ref{fig:cvr-box-sml}.  
The thick lines shows the
median gap, the cross the mean.  Box edges are first and third
quartiles, while circles show gaps that are outside the quartiles
by more than $1.5$ times the inter-quartile range.
Methods for which the maximum gap is greater than the range of the graph
are marked on the right margin with ``$>$''.
}
\label{fig:abv-blw-box-neut}
\end{figure}

Balance is the third desideratum for intervals.  Interval balance
for the neutral scenario is shown in Figure~\ref{fig:abv-blw-box-neut}.
The MLE normal method tends to overstate recall (lower gap larger
than upper), and the adjusted
normal methods to understate it, because they respectively
understate and overstate unretrieved prevalence.
The uniform prior, though unbiased (Figure~\ref{fig:cvr-box-sml}),  
tends to understatement; the prior on the unretrieved segment is sometimes
too strong.
The most conservative prior often overstates
recall, while the half prior is more
balanced.  The legal
and small scenarios (not shown for space) give similar 
results, with the half-prior
beta-binomial being consistently the best balanced.

\subsubsection{Interval width}
\label{sect:width}

\newcommand{\duocol}[1]{\multicolumn{2}{c}{#1}}
\begin{table}
\tbl{Mean width of 95\% recall interval for different interval
methods on the three scenarios, along with the mean coverage of
these methods.
\label{tbl:mean-width}}{
\begin{tabular}{l rr rr rr}
\toprule
 \multirow{2}{*}{Method}
 & \duocol{Neutral} & \duocol{Legal} & \duocol{Small} \\
 \cmidrule(r){2-3} \cmidrule(r){4-5} \cmidrule(r){6-7}
 & width & cvr & width & cvr & width & cvr \\
\midrule
 Naive Binomial & 0.05 & 0.90 & 0.13 & 0.70 & 0.16 & 0.93 \\ 
  Normal MLE & 0.05 & 0.94 & 0.23 & 0.86 & 0.14 & 0.93 \\ 
  Normal Laplace & 0.05 & 0.95 & 0.27 & 0.90 & 0.14 & 0.95 \\ 
  Normal Agresti & 0.05 & 0.95 & 0.24 & 0.83 & 0.14 & 0.93 \\ 
  Koopman & 0.05 & 0.95 & 0.27 & 0.95 & 0.16 & 0.97 \\ 
  Beta, Jeffreys & 0.05 & 0.95 & 0.28 & 0.95 & 0.13 & 0.92 \\ 
  BetaBin, Uniform & 0.05 & 0.95 & 0.26 & 0.95 & 0.14 & 0.95 \\ 
  BetaBin, MCP & 0.05 & 0.95 & 0.28 & 0.95 & 0.14 & 0.95 \\ 
  BetaBin, Half & 0.05 & 0.95 & 0.28 & 0.95 & 0.14 & 0.95 \\ 
  
\bottomrule
\end{tabular}
}
\end{table}

Mean interval width is given in
Table~\ref{tbl:mean-width}, along with mean coverage.
The widths of biased
intervals tend in the same direction as the interval's
bias, though not universally: the naive binomial method
provides an over-wide but under-covering (because highly
imbalanced) interval on the small scenario.
Unbiased intervals have similar widths; balance and 
consistency should be used to choose between them.

\subsubsection{Discussion}
\label{sect:eval-discuss}

\begin{table}[t]
\tbl{Root mean squared error from nominal coverage
for interval estimators on the three different scenarios.
\label{tbl:cvr-rmse}}{
\begin{tabular}{l c rrr c r}
\toprule
\multirow{2}{*}{Method} && \multicolumn{3}{c}{Scenario} 
                         && \multirow{2}{*}{Mean} \\
\cmidrule{3-5}
  && Neutral & Legal & Small &&   \\ 
  \midrule
Naive Binomial  && 0.183 & 0.395 & 0.117 &&  0.231 \\ 
  Normal MLE  && 0.043 & 0.197 & 0.081 &&  0.107 \\ 
  Normal Laplace  && 0.012 & 0.092 & 0.022 &&  0.042 \\ 
  Normal Agresti  && 0.022 & 0.193 & 0.056 &&  0.090 \\ 
  Koopman  && 0.007 & 0.010 & 0.024 &&  0.014 \\ 
  Beta, Jeffreys  && 0.008 & 0.014 & 0.037 &&  0.020 \\ 
  BetaBin, Uniform  && 0.007 & 0.013 & 0.009 &&  0.010 \\ 
  BetaBin, MCP  && 0.009 & 0.015 & 0.011 &&  0.012 \\ 
  BetaBin, Half  && 0.008 & 0.014 & 0.010 &&  0.010 \\ 
   \bottomrule

\end{tabular}
}
\end{table}

We summarize the bias and consistency of
the examined interval methods in Table~\ref{tbl:cvr-rmse},
as the root mean squared error (RMSE) between actual and nominal
coverage.
The naive binomial method is biased and highly inconsistent for
all the scenarios considered.  The normal methods
are less biased, with the Laplace adjustment offering
reasonable average performance, but are highly unstable,
with coverage often falling far below the nominal level
(Figures~\ref{fig:cvr-box-neut}, \ref{fig:cvr-box-lgl},
and \ref{fig:cvr-box-sml}).
These interval methods should be avoided.
The Koopman ratio-of-binomial method, and the beta 
and beta-binomial posterior methods, are more reliable.
Where the sample is a large proportion of the population,
however, as in the small scenario, the Koopman method
leads to over-coverage, and the Jeffreys to under-coverage.
The uniform, most conservative, and half priors to
the beta-binomial posterior offer similar unbiasedness
and consistency, but the former two are less balanced,
the uniform understating recall, and the most
conservative overstating it (Figure~\ref{fig:abv-blw-box-neut}).
Though there is not a great difference between the performance
of the uniform and the half prior to the beta-binomial,
the latter is slightly more balanced, and is therefore the 
recommended interval.  

\begin{figure}
\centering
\includegraphics{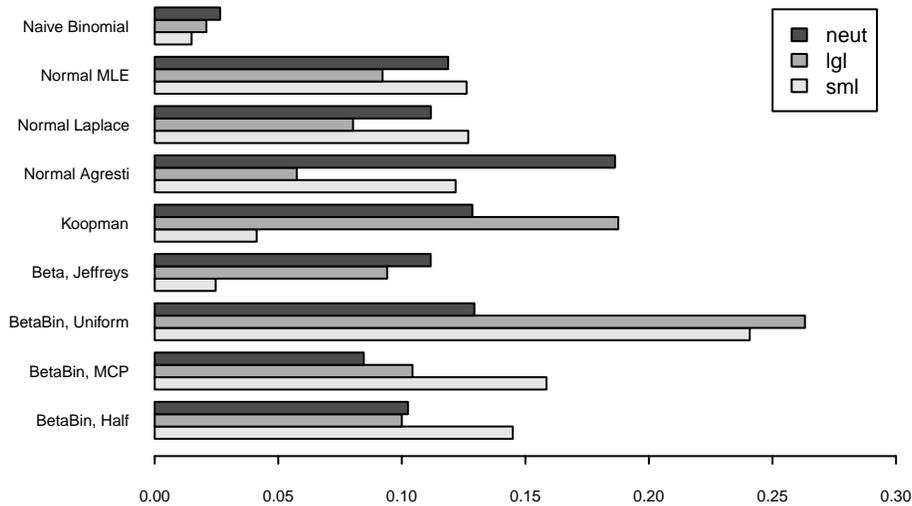}
\caption{Proportion of scenario realizations for which each interval method
gives coverage closest to the nominal level, for each of
the evaluation scenarios.  Tied coverage (to the fidelity of
the simulation, which is $1{,}000$ simulated samples per
realization) receives split counts.}
\label{fig:closest-cvr}
\end{figure}

Do methods with better mean accuracy perform better in the most cases?
Figure~\ref{fig:closest-cvr} shows that the answer is no.
Though the Koopman and posterior methods have 
as little as a tenth of the RMSE of the normal methods, they
outperform only on a moderate majority of individual realizations.
And although the most-conservative prior and 
half prior beta-binomial
have similar RMSE to the uniform prior, the uniform prior
comes closest to nominal coverage in a plurality of
realizations for all three scenarios.
The weaker methods underperform not by being slightly less accurate
on most cases, but by being much less accurate in particular
circumstances.  The evaluator, though, cannot tell from
the sample precisely which circumstance the population fits
into; therefore, the overall most accurate interval
is advisable.

\subsection{Sampling design and expected intervals}
\label{sect:design}

The previous sections have discussed the retrospective analysis
of recall confidence intervals.  The experimental designer or
system auditor, however, is faced with several prospective
questions.  How wide a confidence interval is an analysis likely
to produce?  How large a sample size is needed to reduce
intervals to reasonable width?  And what division
of samples between retrieved and unretrieved segments minimizes
width?

\begin{figure}
\centering
\includegraphics{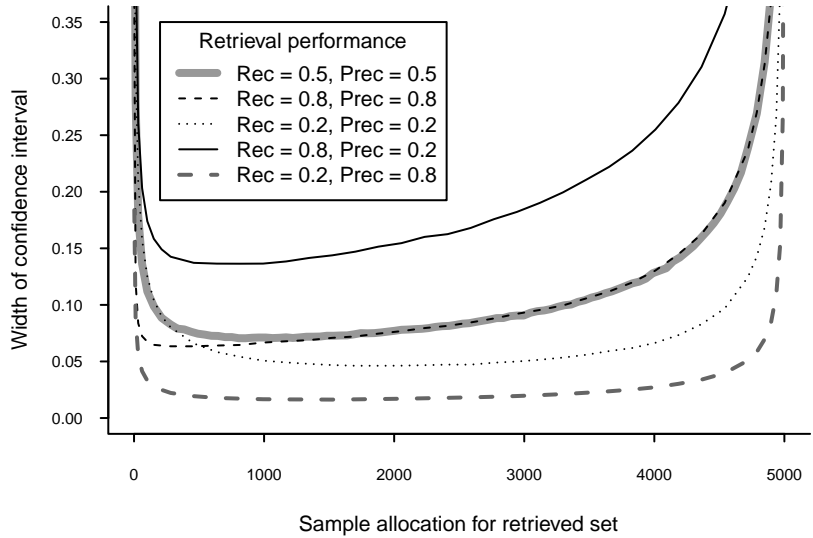}
\caption{The width of confidence intervals, using the
beta-binomial posterior with half prior, for retrievals
of different levels of effectiveness.  A sample of $n$
is drawn from the retrieved segment and $5{,}000 - n$ documents
from the unretrieved segment, from a population of 
$5{,}000{,}000$
documents, and a retrieved segment size of 
$500{,}000$.
}
\label{fig:ci-wd-perf}
\end{figure}

Interval width depends on sample size; division of sample
amongst segments; corpus yield; recall; retrieval size; 
and (though only marginally) population size.  
Sample size and sample allocation are under the 
evaluator's control; retrieval and population size are
known at design time; but 
yield and recall are unknown prior to sampling.
The sensitivity of allocation strategies to retrieval performance
is examined in Figure~\ref{fig:ci-wd-perf}.  Interval width is
clearly affected by retrieval quality: low recall and high precision
leads to narrow intervals, since prevalence is high and sampling
variance low in both retrieved and unretrieved segments. 
Here, width is not sensitive to allocation.  In contrast,
low precision and high recall causes wide intervals, and favors
an allocation of samples towards the unretrieved segment.   
A 20\%:80\% sample allocation to the retrieved and unretrieved
segments gives reasonably close to minimal interval width across the
scenarios, but would obstruct the accurate estimation of precision.

\begin{figure}
\centering
\includegraphics{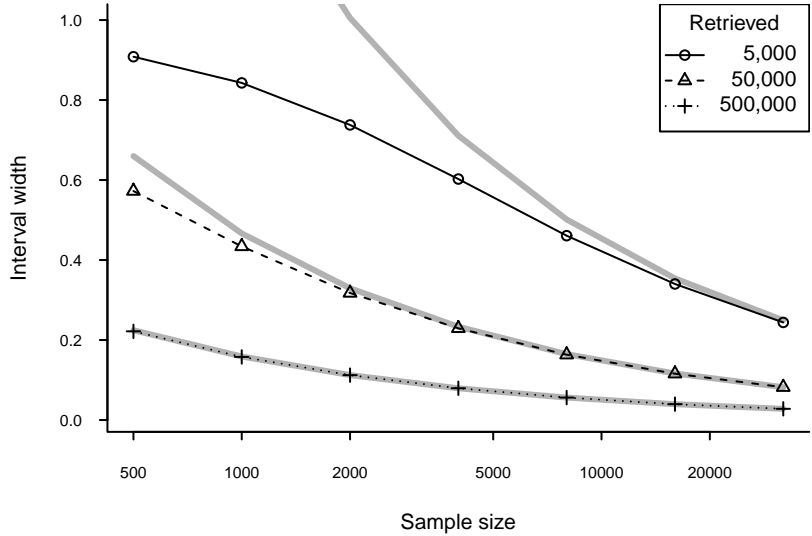}
\caption{Width of recall confidence interval for varying
sample sizes and different retrieved segment sizes.  The
population size is $\num{5e6}$, and the retrieval has recall
and precision both $0.5$.  Interval widths are shown for the
optimal allocation of samples and assessments to the retrieved
and unretrieved segments.  The black lines show the beta-binomial
posterior with half prior.  The grey lines show the
intervals given by the MLE normal method.}
\label{fig:ci-wd-min-int}
\end{figure}

Allocation aside, the chief variable at the
evaluator's control is sample
size.  Figure~\ref{fig:ci-wd-min-int} shows interval width
as a function of sample size 
for the half-prior beta-binomial 
and MLE normal methods, assuming
an ideal sample allocation between segments.  The normal method shows the
standard $1/\sqrt{n}$ reduction in interval width with
an $n$-fold increase in sample size, but 
gives overly wide intervals for low prevalence
and small samples, producing
intervals wider than $1.0$ in some cases.  The beta-binomial
agrees closely with the normal for high prevalence and large
samples, but gives smaller intervals for low prevalence and
small samples.  Consequently, the decrease in interval width
with sample size is slower than $1/\sqrt{n}$ in such circumstances.

\subsection{Intervals for TREC Legal Track}
\label{sect:trec-int}

We conclude the experimental section by calculating
confidence intervals for the Interactive Task of
the TREC 2008 and TREC 2009 Legal Track.  The
task was evaluated
using stratified sampling (Section~\ref{sect:strat}),
with strata defined by the intersection of participant
team retrievals; $n$ teams define $2 ^ n$ strata,
some of which may be empty.  Different strata received
different sampling rates; in particular, the bottom stratum,
of documents retrieved by no system, was sampled only sparsely
\cite{legal08:trec}.  The recall confidence interval for the
official results was calculated using the Normal-MLE
method (Section~\ref{sect:normal-mle}), though with a 
propagation of error expression that omits 
yield covariance (Section~\ref{sect:prop-err}).

\begin{table}
\tbl{Recall estimates with 95\% confidence intervals for the 
Interactive Task of the TREC 2008 Legal Track.  The beta-binomial 
interval uses the $\alpha = \beta = 0.5$ as prior.  The normal interval
use MLE variance estimates, without and with correction 
for the correlation between system and corpus yield estimates.
\label{tbl:int-rec-08}
}{
\begin{tabular}{ll r c rr c rr c rr}
\toprule
\multirow{2}{*}{Topic} & \multirow{2}{*}{Team} & \multirow{2}{*}{Recall} && \multicolumn{2}{c}{CI (BBin, 0.5)} && \multicolumn{2}{c}{CI (Norm, uncorr.)} && \multicolumn{2}{c}{CI (Norm, corr.)}\\ 
\cmidrule{5-6} \cmidrule{8-9} \cmidrule{11-12}
 & & && 2.5\% & 97.5\% && 2.5\% & 97.5\% && 2.5\% & 97.5\% \\ 
\midrule
t102 & Ad Hoc Pool & 0.314 && 0.273 & 0.360 && 0.266 & 0.362 && 0.271 & 0.358 \\
     &   Clearwell & 0.016 && 0.014 & 0.018 && 0.014 & 0.018 && 0.014 & 0.018 \\
     &  Pittsburgh & 0.007 && 0.006 & 0.008 && 0.006 & 0.008 && 0.006 & 0.008 \\[0.5em]
t103 &          H5 & 0.624 && 0.580 & 0.665 && 0.579 & 0.668 && 0.581 & 0.666 \\
     & Ad Hoc Pool & 0.403 && 0.374 & 0.430 && 0.371 & 0.434 && 0.374 & 0.431 \\
     &   Clearwell & 0.158 && 0.146 & 0.169 && 0.146 & 0.169 && 0.146 & 0.169 \\
     &     Buffalo & 0.061 && 0.056 & 0.066 && 0.056 & 0.066 && 0.056 & 0.066 \\
     &  Pittsburgh & 0.026 && 0.024 & 0.028 && 0.024 & 0.029 && 0.024 & 0.029 \\[0.5em]
t104 & Ad Hoc Pool & 0.345 && 0.185 & 0.577 && 0.111 & 0.580 && 0.143 & 0.548 \\
     &   Clearwell & 0.003 && 0.002 & 0.005 && 0.001 & 0.004 && 0.001 & 0.004 \\
\bottomrule 
\end{tabular} 
}
\end{table}

\begin{table}
\tbl{Recall and interval estimates for the Interactive Task
of the TREC 2009 Legal Track.  The MLE normal and Laplace methods
correct for yield estimate correlation.
\label{tbl:int-rec-09}
}{

\begin{tabular}{ll r c rr c rr c rr}
\toprule
\multirow{2}{*}{Topic} & \multirow{2}{*}{Team} & \multirow{2}{*}{Recall} && \multicolumn{2}{c}{CI (BBin, 0.5)} && \multicolumn{2}{c}{CI (Norm, corr.)} && \multicolumn{2}{c}{CI (Laplace)}\\ 
\cmidrule{5-6} \cmidrule{8-9} \cmidrule{11-12}
 & & && 2.5\% & 97.5\% && 2.5\% & 97.5\% && 2.5\% & 97.5\% \\ 
\midrule
t201 &        UW & 0.778 && 0.451 & 0.928 && 0.485 & 1.070 && 0.351 & 0.916 \\
     &        CS & 0.489 && 0.287 & 0.592 && 0.304 & 0.673 && 0.226 & 0.593 \\
     &        CB & 0.204 && 0.119 & 0.250 && 0.126 & 0.282 && 0.093 & 0.247 \\
     &        UP & 0.167 && 0.097 & 0.208 && 0.102 & 0.231 && 0.076 & 0.205 \\[0.5em]
t202 &        UW & 0.673 && 0.523 & 0.768 && 0.542 & 0.803 && 0.501 & 0.771 \\
     &        CS & 0.579 && 0.451 & 0.662 && 0.467 & 0.692 && 0.432 & 0.664 \\[0.5em]
t203 &        UW & 0.865 && 0.623 & 0.881 && 0.832 & 0.897 && 0.552 & 0.940 \\
     &        UB & 0.592 && 0.427 & 0.621 && 0.552 & 0.631 && 0.378 & 0.652 \\
     & ZL-NoCull & 0.175 && 0.123 & 0.185 && 0.157 & 0.192 && 0.105 & 0.185 \\
     &   ZL-Cull & 0.029 && 0.019 & 0.034 && 0.023 & 0.035 && 0.016 & 0.033 \\[0.5em]
t204 &        H5 & 0.762 && 0.572 & 0.882 && 0.595 & 0.930 && 0.538 & 0.874 \\
     &        UD & 0.305 && 0.229 & 0.360 && 0.235 & 0.374 && 0.216 & 0.358 \\
     &        CB & 0.198 && 0.149 & 0.240 && 0.151 & 0.246 && 0.140 & 0.238 \\[0.5em]
t205 &        CS & 0.673 && 0.592 & 0.738 && 0.599 & 0.747 && 0.587 & 0.737 \\
     &        EQ & 0.463 && 0.404 & 0.509 && 0.409 & 0.516 && 0.399 & 0.505 \\
     &        IN & 0.292 && 0.253 & 0.327 && 0.254 & 0.329 && 0.250 & 0.325 \\[0.5em]
t206 &   CB-High & 0.076 && 0.052 & 0.112 && 0.046 & 0.106 && 0.049 & 0.109 \\
     &        LO & 0.042 && 0.025 & 0.069 && 0.020 & 0.063 && 0.023 & 0.067 \\
     &    CB-Mid & 0.011 && 0.007 & 0.015 && 0.007 & 0.015 && 0.007 & 0.014 \\
     &    CB-Low & 0.009 && 0.006 & 0.013 && 0.006 & 0.013 && 0.005 & 0.012 \\[0.5em]
t207 &        CB & 0.768 && 0.698 & 0.798 && 0.723 & 0.813 && 0.689 & 0.802 \\
     &        UW & 0.761 && 0.688 & 0.791 && 0.714 & 0.807 && 0.677 & 0.791 \\
     &        LO & 0.538 && 0.489 & 0.566 && 0.503 & 0.573 && 0.484 & 0.569 \\
     &        EQ & 0.483 && 0.437 & 0.507 && 0.451 & 0.515 && 0.430 & 0.507 \\
\bottomrule 
\end{tabular} 
}
\end{table}

Table~\ref{tbl:int-rec-08} compares the normal and beta-binomial
confidence intervals for TREC 2008.  The ``Ad Hoc Pool'' team was
a deep pseudo-run made up by pooling the automated runs from a
parallel task.  The normal intervals use
the MLE estimate of variance, and are calculated without and with
correction for the correlation between team and corpus yield
estimates (Section~\ref{sect:prop-err}); the uncorrected
interval estimate is as reported in the official results 
\cite{legal08:trec}.  The correlation-corrected interval differs
substantially from the uncorrected only for the Ad Hoc Pool pseudo-team
on Topic~104, since only then does it display the necessary
combination of a high-recall, low-precision ($0.023$) run.
It is also only on Topic~104 that there is much difference
between the corrected MLE normal and
the beta-binomial intervals, since the unretrieved segment 
sample prevalence is not small for the other two topics.

The normal and beta-binomial intervals for TREC 2009 are shown
in Table~\ref{tbl:int-rec-09}.  Correlation correction is applied
for the normal version.  The uncorrected intervals (not shown)
are marginally narrower in most cases, and sharply narrower
for a few.  
In contrast to the previous
year, the normal and beta-binomial intervals generally disagree.  
The mean absolute
difference in width is $13\%$, with the beta-binomial
being narrower for all topics save Topic 203.  The
lower bound of the beta-binomial generally falls below that
of the normal, 
and is wider beneath the point estimate than above it.  
The cause is low or zero sample prevalence in the bottom
stratum, to which the beta-binomial method correctly gives
a wider upper than lower bound.  The normal method 
assigns an impossible upper-bound recall of $1.07$, even
though the team fails to return several known relevant documents.
For comparison, the final two columns of Table~\ref{tbl:int-rec-09} 
show the Laplace adjustment to the normal interval 
(Section~\ref{sect:normal-adj}).  The adjusted interval
is generally left-skewed, like the beta-binomial.  
The intervals, however, are wider, sometimes substantially so, 
than the beta-binomial.

\section{Conclusion}
\label{sect:concl}

Several recall confidence interval methods have been considered 
in this article.  The naive binomial method, in which recall is
treated
as a simple binomial proportion on the relevant documents, assumes
equal sampling of retrieved and unretrieved segments, which is
not generally the case in retrieval evaluation.  The normal
approximation, with propagation of per-segment yield sampling error,
understates variance on the unretrieved segment, and is strongly
biased towards undercoverage. The Laplace and Agresti-Coull
adjustments only partially correct this bias, and generalize
poorly to stratified sampling.  The Koopman analytical ratio-of-binomial
estimator gives much better results, as does the
ratio-of-binomials posterior to the Jeffreys prior. Both, however,
are inaccurate when a large proportion of the population is sampled.

The most accurate approximate interval
infers a beta-binomial posterior for each segment or stratum, 
and samples from this posterior
to generate a distribution over recall.
The interval is unbiased, with mean coverage at the nominal $1 - \alpha$
level, and is stable, with quartiles clustering
to within a couple of percentage points of the nominal level.
The greatest stability and balance is given by a 
prior that sets the hyperparameters $\alpha$ and $\beta$ to $0.5$;
this is more accurate than either the uniform or the
(more theoretically grounded) most-conservative prior.
(Our results are summarized in the final column of Table~\ref{tbl:rec-ci},
on page~\pageref{tbl:rec-ci}.)

We have provided guidance on experimental design.  
Interval width is generally minimized
by assigning most assessments to the unretrieved segment,
particularly for high-recall retrievals, though such an allocation
will be at the cost of wider intervals on precision.
Assuming optimal allocation between segments, interval width 
under the beta-binomial prior method decreases by the usual 
square root of sample size only once the width of the interval is well below
$0.5$; before that, decrease in interval width is slower.

Finally, we have used the half-prior 
beta-binomial posterior method
recommended in this article to produce interval
estimates on the TREC 2008 and TREC 2009 Legal Track, Interactive
Task participants, and compared them with the unadjusted and
adjusted normal methods.  In some cases, the normal methods
give similar estimates to the beta-binomial; in others,
marked differences exist, particularly where sample prevalence in
the unretrieved segment is low.

\subsection{Future work}
\label{sect:future}

This article has considered two-tailed confidence
intervals.  One-tailed intervals, particularly those setting
a lower bound on performance, are also of interest.  
Achieving accurate one-tailed
intervals can be more demanding than for two-tailed intervals, 
since under-coverage
at one end cannot be compensated by over-coverage at the other.
For instance, \citeN{cai05:jspi} finds that though the two-tailed
Wilson interval on the binomial is highly accurate, the one-tailed
interval is systematically biased.  
\citeN{lk09:jos} find that the beta
posterior with Jeffreys prior gives an accurate one-tailed
interval on the binomial proportion, and the good balance
observed for the two-tailed half-prior beta-binomial interval
on recall (Figure~\ref{fig:abv-blw-box-neut}) suggests that
it would also provide an accurate one-tailed interval.
This article has also only considered approximate intervals, 
aiming at nominal mean coverage.  In some circumstances,
a guaranteed minimal coverage may be required.  For such
cases, an exact method is needed.

We noted in Section~\ref{sect:rec-dist} that the recall estimator
is biased, and that the bias can be severe and positive, particularly where
the prevalence of the unretrieved segment is low, and the unretrieved
sample size is small.  There is long-standing
literature on unbiased ratio estimators~\cite{hr54:nature,alj08:phd}.
An unbiased estimator of recall is desirable.

It was also noted in Section~\ref{sect:rec-dist} that the sampling
distribution of recall is generally asymmetric.  For the similarly
asymmetric sampling distribution of the binomial proportion,
a two-tailed Wald interval on $\logit{\pi}$ of
$\ln p / q \pm z / \sqrt{npq}$ has been found to give coverage
characteristics similar to the 
Wilson interval~\cite{newc01:tas}, 
as has the one-tailed variant~\cite{lk09:jos}.  Applying
such a transform to the normal approximation interval for recall
might correct its spurious symmetry.
The transformed interval would still mishandle extreme cases, though; 
for instance, when estimated
recall is $0$ or $1$, then the logit-transformed intervals
are $[0,0]$ and $[1,1]$, respectively.

The optimal allocation of samples to retrieved and unretrieved segments
depends upon prevalence in each set (Section~\ref{sect:design}).  
Prevalence is not known in advance, but could be estimated 
during sampling.  Additionally, sampling could finish once
a certain interval width or lower bound is reached.  Such sequential
analysis can lead to biased estimates~\cite{wg86:seqmeth}; 
the degree of this bias would need to be determined.

We have assumed there is no assessor
error.  Such error has a particularly strong
impact on low-prevalence corpora;
even a low false positive rate can drastically inflate estimated
yield.  The error rate can be estimated and corrected by
sampling primary assessments for checking by a more
authoritative assessor (if available)~\cite{wosh10:cikm}, though this 
estimate will have its own sampling error, and the
authoritative assessor might also make
mistakes.  Other approaches might overlap multiple
assessors, to verify dubious assessments and estimate assessor
reliability.  

Only simple and stratified random sampling have been
considered in this article.  A more general approach is
unequal sampling, in which each document is assigned
its own inclusion probability, based on its probability
of relevance and its weight in the evaluation metric.
Unequal sampling has been applied in
the Ad Hoc Task of the TREC Legal Track~\cite{legal07:trec,legal09:trec},
and methods for unequal sampling and point 
estimation in ranked retrieval evaluation have been 
developed~\cite{apy06:sigir,ap08:tr}.  Estimating variance
for unequal sampling schemes requires calculating joint
inclusion probabilities for sampled pairs.  A simple 
sampling method for which joint inclusion probabilities
are known is Sunter sampling~\cite{sunter77:jrss,ssw92:mass}; as 
it happens, the sampling scheme used in the Legal Ad Hoc Task
is equivalent to Sunter sampling.  \citeN{pavlu08:phd} provides
a variance estimator for the average precision metric, though it
understates variance by ignoring a component of variance.
Such variance estimators,
however, rely upon normal approximations to turn into confidence
intervals, and handle zero-prevalence samples no better than
simple random sampling approaches.
Bayesian methods are also more complex to use
with unequal sampling, since we are no longer estimating a
distribution over a single parameter (such as the proportion
relevant in a segment), but require a more complex model linking
probability of relevance to rank and other evidence~\cite{cart07:sigir}.

Intervals for set-based metrics other than recall, such as precision and
F1 score, can be inferred using the same beta-binomial posteriors 
on segment or stratum yield.  The characteristics and
accuracy of these intervals need to be established, and
compared with other interval methods, particularly for the interval
on precision, which (as a simple binomial proportion) has
a plethora of intervals available.

Simple or stratified random sampling requires minimal
assumptions, and produces wide bounds on
yield and recall.  Stronger assumptions can give
tighter bounds, but shift the uncertainty to the validity of
the assumptions.  For instance, \citeN{kkia99:asis} 
propose a capture-recapture model, using the overlap 
between two runs to estimate yield.  In its simplest form,
this requires the unrealistic assumption that the two
runs are independent; but a model that accounts for run
correlation could be developed.  
\citeN{z98:sigir} fits and extrapolates a model of proportion
relevant at depth in ranked retrievals.  Extending to
depth one million a curve that has been observed to depth
one hundred requires some courage, but the approach could be
combined with sampling and other forms of evidence.  As with
simple random sampling, confidence intervals are required
as well as point estimates

Whatever estimation methods are used, and even if assessor
error is excluded, bounds on recall for
large collections with low prevalence and constrained 
assessment budgets will generally be wide, sometimes so
wide as to appear unhelpful to the evaluator.  But this
is no reason to neglect the calculation and reporting
of such bounds; quite the reverse.  Where uncertainty is
high, it is all the more imperative to reveal that fact.
Only once the uncertainty is recognized can the effort be made
to bring it within tolerable limits.

\refsection*{REPRODUCIBLE RESULTS STATEMENT}

An implementation of the recall confidence interval methods,
along with other code and datasets used in this paper,
is available at \url{codalism.com/~wew/w12recci}.

\par

\begin{acks}
The author thanks Doug Oard, Dave Lewis, Eric Slud, 
and Mossaab Bagdouri, and the anonymous reviewers, for their
perceptive suggestions.
\end{acks}

\bibliography{localbib}

\begin{thebibliography}{}

\bibitem[\protect\citeauthoryear{Agresti and Caffo}{Agresti and
  Caffo}{2000}]{ac00:tas}
{\sc Agresti, A.} {\sc and} {\sc Caffo, B.} 2000.
\newblock Simple and effective confidence intervals for proportions and
  differences of proportions result from adding two successes and two failures.
\newblock {\em The American Statistician\/}~{\em 54,\/}~4, 280--288.

\bibitem[\protect\citeauthoryear{Agresti and Coull}{Agresti and
  Coull}{1998}]{ac98:tas}
{\sc Agresti, A.} {\sc and} {\sc Coull, B.~A.} 1998.
\newblock Approximate is better than ``exact'' for interval estimation of
  binomial proportions.
\newblock {\em The American Statistician\/}~{\em 52,\/}~2, 119--126.

\bibitem[\protect\citeauthoryear{Al-Jararha}{Al-Jararha}{2008}]{alj08:phd}
{\sc Al-Jararha, J.} 2008.
\newblock Unbiased ratio estimation for finite populations.
\newblock Ph.D. thesis, Colorado Statue University.

\bibitem[\protect\citeauthoryear{Aslam and Pavlu}{Aslam and
  Pavlu}{2008}]{ap08:tr}
{\sc Aslam, J.} {\sc and} {\sc Pavlu, V.} 2008.
\newblock A practical sampling strategy for efficient retrieval evaluation.
\newblock Tech. rep., Northeastern University.

\bibitem[\protect\citeauthoryear{Aslam, Pavlu, and Yilmaz}{Aslam
  et~al\mbox{.}}{2006}]{apy06:sigir}
{\sc Aslam, J.}, {\sc Pavlu, V.}, {\sc and} {\sc Yilmaz, E.} 2006.
\newblock A statistical method for system evaluation using incomplete
  judgments.
\newblock In {\em Proc. 29th Annual International {ACM} {SIGIR} Conference on
  Research and Development in Information Retrieval}, {S.~Dumais},
  {E.~Efthimiadis}, {D.~Hawking}, {and} {K.~J{\"{a}}rvelin}, Eds. Seattle,
  Washington, USA, 541--548.

\bibitem[\protect\citeauthoryear{Berger, Bernardo, and Sun}{Berger
  et~al\mbox{.}}{2008}]{bbs08:tr}
{\sc Berger, J.~O.}, {\sc Bernardo, J.~M.}, {\sc and} {\sc Sun, D.} 2008.
\newblock Objective priors for discrete parameter spaces.
\newblock Tech. rep., Duke University.

\bibitem[\protect\citeauthoryear{Bolstad}{Bolstad}{2007}]{bolstad07:byst}
{\sc Bolstad, W.~M.} 2007.
\newblock {\em Introduction to Bayesian Statistics}.
\newblock John Wiley \& Sons.

\bibitem[\protect\citeauthoryear{Brown, Cai, and Das{G}upta}{Brown
  et~al\mbox{.}}{2001}]{bcd01:statsci}
{\sc Brown, L.~D.}, {\sc Cai, T.~T.}, {\sc and} {\sc Das{G}upta, A.} 2001.
\newblock Interval estimation for a binomial proportion.
\newblock {\em Statistical Science\/}~{\em 18,\/}~2, 101--133.

\bibitem[\protect\citeauthoryear{Buckland}{Buckland}{1984}]{buck84:biom}
{\sc Buckland, S.~T.} 1984.
\newblock {M}onte {C}arlo confidence intervals.
\newblock {\em Biometrics\/}~{\em 40,\/}~3, 811--817.

\bibitem[\protect\citeauthoryear{Cai}{Cai}{2005}]{cai05:jspi}
{\sc Cai, T.~T.} 2005.
\newblock One-sided confidence intervals in discrete distributions.
\newblock {\em Journal of Statistical Planning and Inference\/}~{\em 131,\/}~1,
  63--88.

\bibitem[\protect\citeauthoryear{Carterette}{Carterette}{2007}]{cart07:sigir}
{\sc Carterette, B.} 2007.
\newblock Robust test collections for retrieval evaluation.
\newblock In {\em Proc. 30th Annual International {ACM} {SIGIR} Conference on
  Research and Development in Information Retrieval}, {C.~L.~A. Clarke},
  {N.~Fuhr}, {N.~Kando}, {W.~Kraaij}, {and} {A.~de~Vries}, Eds. Amsterdam, the
  Netherlands, 55--62.

\bibitem[\protect\citeauthoryear{Chen and Shao}{Chen and
  Shao}{1999}]{cs99:jcgs}
{\sc Chen, M.-H.} {\sc and} {\sc Shao, Q.-M.} 1999.
\newblock {M}onte {C}arlo estimation of {B}ayesian credible and {HPD}
  intervals.
\newblock {\em Journal of Computational and Graphical Statistics\/}~{\em
  8,\/}~1, 69--92.

\bibitem[\protect\citeauthoryear{Cheng}{Cheng}{1978}]{cheng78:cacm}
{\sc Cheng, R. C.~H.} 1978.
\newblock Generating beta variates with nonintegral shape parameters.
\newblock {\em Communications of the ACM\/}~{\em 21,\/}~4, 317--322.

\bibitem[\protect\citeauthoryear{Clopper and Pearson}{Clopper and
  Pearson}{1934}]{cp34:biom}
{\sc Clopper, C.~J.} {\sc and} {\sc Pearson, E.~S.} 1934.
\newblock The use of confidence or fiducial limits illustrated in the case of
  the binomial.
\newblock {\em Biometrika\/}~{\em 26,\/}~4, 404--413.

\bibitem[\protect\citeauthoryear{Cochran}{Cochran}{1977}]{cochran77}
{\sc Cochran, W.~G.} 1977.
\newblock {\em Sampling Techniques\/} 3rd Ed.
\newblock John Wiley \& Sons.

\bibitem[\protect\citeauthoryear{Dutka}{Dutka}{1984}]{dutka84:ahes}
{\sc Dutka, J.} 1984.
\newblock The early history of the hypergeometric function.
\newblock {\em Archive for History of Exact Sciences\/}~{\em 31,\/}~1, 15--34.

\bibitem[\protect\citeauthoryear{Dyer and Chiou}{Dyer and
  Chiou}{1984}]{dc84:cstm}
{\sc Dyer, D.} {\sc and} {\sc Chiou, P.} 1984.
\newblock An information-theoretic approach to incorporating prior information
  in binomial sampling.
\newblock {\em Communications in Statistics: Theory and Methods\/}~{\em
  13,\/}~17, 2051--2083.

\bibitem[\protect\citeauthoryear{Dyer and Pierce}{Dyer and
  Pierce}{1993}]{dp93:cstm}
{\sc Dyer, D.} {\sc and} {\sc Pierce, R.~L.} 1993.
\newblock On the choice of the prior distribution in hypergeometric sampling.
\newblock {\em Communications in Statistics: Theory and Methods\/}~{\em
  22,\/}~8, 2125--2146.

\bibitem[\protect\citeauthoryear{Feller}{Feller}{1945}]{feller45:ams}
{\sc Feller, W.} 1945.
\newblock On the normal approximation to the binomial distribution.
\newblock {\em The Annals of Mathematical Statistics\/}~{\em 16,\/}~4,
  319--329.

\bibitem[\protect\citeauthoryear{Gelman, Carlin, Stern, and Rubin}{Gelman
  et~al\mbox{.}}{2004}]{gcsr04:bda}
{\sc Gelman, A.}, {\sc Carlin, J.~B.}, {\sc Stern, H.~S.}, {\sc and} {\sc
  Rubin, D.~B.} 2004.
\newblock {\em Bayesian Data Analysis\/} 2nd Ed.
\newblock Chapman and Hall/CRC.

\bibitem[\protect\citeauthoryear{Greenland}{Greenland}{2001}]{greenland01:tas}
{\sc Greenland, S.} 2001.
\newblock {A}gresti, {A}., and {C}affo, {B}., ``{S}imple and effective
  confidence intervals for proportions and differences of proportions result
  from adding two successes and two failures,'' {T}he {A}merican
  {S}tatistician, 54, 280--288: {C}omment by {G}reenland and reply.
\newblock {\em The American Statistician\/}~{\em 55,\/}~2, 172.

\bibitem[\protect\citeauthoryear{Guenther}{Guenther}{1971}]{guen71:tas}
{\sc Guenther, W.~C.} 1971.
\newblock Unbiased confidence intervals.
\newblock {\em The American Statistician\/}~{\em 25,\/}~1, 51--53.

\bibitem[\protect\citeauthoryear{Hartley and Ross}{Hartley and
  Ross}{1954}]{hr54:nature}
{\sc Hartley, H.~O.} {\sc and} {\sc Ross, A.} 1954.
\newblock Unbiased ratio estimators.
\newblock {\em Nature\/}~{\em 174}, 270--271.

\bibitem[\protect\citeauthoryear{Hedin, Tomlinson, Baron, and Oard}{Hedin
  et~al\mbox{.}}{2009}]{legal09:trec}
{\sc Hedin, B.}, {\sc Tomlinson, S.}, {\sc Baron, J.~R.}, {\sc and} {\sc Oard,
  D.~W.} 2009.
\newblock Overview of the {TREC} 2009 legal track.
\newblock In {\em Proc. 18th Text {RE}trieval Conference}, {E.~Voorhees} {and}
  {L.~P. Buckland}, Eds. Gaithersburg, Maryland, USA, 1:4:1--40.
\newblock {NIST} Special Publication 500-278.

\bibitem[\protect\citeauthoryear{Jeffreys}{Jeffreys}{1946}]{jeff46:prs}
{\sc Jeffreys, H.} 1946.
\newblock An invariant form for the prior probability in estimation problems.
\newblock {\em Proc. Royal Society of London. Series A, Mathematical and
  Physical Sciences\/}~{\em 186,\/}~1007, 453--461.

\bibitem[\protect\citeauthoryear{Kantor, Kim, Ibraev, and Atasoy}{Kantor
  et~al\mbox{.}}{1999}]{kkia99:asis}
{\sc Kantor, P.}, {\sc Kim, M.-H.}, {\sc Ibraev, U.}, {\sc and} {\sc Atasoy,
  K.} 1999.
\newblock Estimating the number of relevant documents in enormous collections.
\newblock In {\em Proc. ASIS Annual Meeting}. 507--514.

\bibitem[\protect\citeauthoryear{Koopman}{Koopman}{1984}]{koopman84:biom}
{\sc Koopman, P. A.~R.} 1984.
\newblock Confidence intervals for the ratio of two binomial proportions.
\newblock {\em Biometrics\/}~{\em 40,\/}~2, 513--517.

\bibitem[\protect\citeauthoryear{Lehmann and Romano}{Lehmann and
  Romano}{2005}]{lr05:stathyp}
{\sc Lehmann, E.~L.} {\sc and} {\sc Romano, J.~P.} 2005.
\newblock {\em Testing Statistical Hypotheses\/} 3rd Ed.
\newblock Springer.

\bibitem[\protect\citeauthoryear{Liu and Kott}{Liu and Kott}{2009}]{lk09:jos}
{\sc Liu, Y.~K.} {\sc and} {\sc Kott, P.~S.} 2009.
\newblock Evaluating alternative one-sided coverage intervals for a proportion.
\newblock {\em Journal of Official Statistics\/}~{\em 25,\/}~4, 569--588.

\bibitem[\protect\citeauthoryear{Newcombe}{Newcombe}{2001}]{newc01:tas}
{\sc Newcombe, R.~G.} 2001.
\newblock Logit confidence intervals and the inverse sinh transformation.
\newblock {\em The American Statistician\/}~{\em 55,\/}~3, 200--202.

\bibitem[\protect\citeauthoryear{Neyman}{Neyman}{1935}]{neyman35:ams}
{\sc Neyman, J.} 1935.
\newblock On the problem of confidence intervals.
\newblock {\em The Annals of Mathematical Statistics\/}~{\em 6,\/}~3, 111--116.

\bibitem[\protect\citeauthoryear{Nicholson}{Nicholson}{1956}]{nich56:ams}
{\sc Nicholson, W.~L.} 1956.
\newblock On the normal approximation to the hypergeometric distribution.
\newblock {\em The Annals of Mathematical Statistics\/}~{\em 27,\/}~2,
  471--483.

\bibitem[\protect\citeauthoryear{Oard, Hedin, Tomlinson, and Baron}{Oard
  et~al\mbox{.}}{2008}]{legal08:trec}
{\sc Oard, D.~W.}, {\sc Hedin, B.}, {\sc Tomlinson, S.}, {\sc and} {\sc Baron,
  J.~R.} 2008.
\newblock Overview of the {TREC} 2008 legal track.
\newblock In {\em Proc. 17th Text {RE}trieval Conference}, {E.~Voorhees} {and}
  {L.~P. Buckland}, Eds. Gaithersburg, Maryland, USA, 3:1--45.
\newblock {NIST} Special Publication 500-277.

\bibitem[\protect\citeauthoryear{Pavlu}{Pavlu}{2008}]{pavlu08:phd}
{\sc Pavlu, V.} 2008.
\newblock Large scale {IR} evaluation.
\newblock Ph.D. thesis, Northeastern University.

\bibitem[\protect\citeauthoryear{Roitblat, Kershaw, and Oot}{Roitblat
  et~al\mbox{.}}{2010}]{rko10:jasist}
{\sc Roitblat, H.~L.}, {\sc Kershaw, A.}, {\sc and} {\sc Oot, P.} 2010.
\newblock Document categorization in legal electronic discovery: computer
  classification vs. manual review.
\newblock {\em Journal of the American Society for Information Science and
  Technology\/}~{\em 61,\/}~1, 70--80.

\bibitem[\protect\citeauthoryear{S{\"{a}}rndal, Swensson, and
  Wretman}{S{\"{a}}rndal et~al\mbox{.}}{1992}]{ssw92:mass}
{\sc S{\"{a}}rndal, C.-E.}, {\sc Swensson, B.}, {\sc and} {\sc Wretman, J.}
  1992.
\newblock {\em Model assisted survey sampling}.
\newblock Springer-Verlag.

\bibitem[\protect\citeauthoryear{Simel, Samsa, and Matchar}{Simel
  et~al\mbox{.}}{1991}]{ssm91:jclep}
{\sc Simel, D.~L.}, {\sc Samsa, G.~P.}, {\sc and} {\sc Matchar, D.~B.} 1991.
\newblock Likelihood ratios with confidence: sample size estimation for
  diagnostic test studies.
\newblock {\em Journal of Clinical Epidemiology\/}~{\em 44,\/}~8, 763--770.

\bibitem[\protect\citeauthoryear{Smithson}{Smithson}{2002}]{smithson02:ci}
{\sc Smithson, M.} 2002.
\newblock {\em Confidence intervals}.
\newblock Sage Publications.

\bibitem[\protect\citeauthoryear{Sunter}{Sunter}{1977}]{sunter77:jrss}
{\sc Sunter, A.~B.} 1977.
\newblock List sequential sampling with equal or unequal probabilities without
  replacement.
\newblock {\em Journal of the Royal Statistical Society. Series C (Applied
  Statistics)\/}~{\em 26,\/}~3, 261--268.

\bibitem[\protect\citeauthoryear{Taylor}{Taylor}{1997}]{taylor97:erranl}
{\sc Taylor, J.~R.} 1997.
\newblock {\em Introduction to error analysis\/} 2nd Ed.
\newblock University Science Books.

\bibitem[\protect\citeauthoryear{Thompson}{Thompson}{2002}]{tho02}
{\sc Thompson, S.~K.} 2002.
\newblock {\em Sampling\/} 2nd Ed.
\newblock John Wiley \& Sons, New York.

\bibitem[\protect\citeauthoryear{Tomlinson, Oard, Baron, and
  Thompson}{Tomlinson et~al\mbox{.}}{2007}]{legal07:trec}
{\sc Tomlinson, S.}, {\sc Oard, D.~W.}, {\sc Baron, J.~R.}, {\sc and} {\sc
  Thompson, P.} 2007.
\newblock Overview of the {TREC} 2007 legal track.
\newblock In {\em Proc. 16th Text {RE}trieval Conference}, {E.~Voorhees} {and}
  {L.~P. Buckland}, Eds. Gaithersburg, Maryland, USA, 5:1--34.
\newblock {NIST} Special Publication 500-274.

\bibitem[\protect\citeauthoryear{Webber, Oard, Scholer, and Hedin}{Webber
  et~al\mbox{.}}{2010}]{wosh10:cikm}
{\sc Webber, W.}, {\sc Oard, D.~W.}, {\sc Scholer, F.}, {\sc and} {\sc Hedin,
  B.} 2010.
\newblock Assessor error in stratified evaluation.
\newblock In {\em Proc. 19th ACM International Conference on Information and
  Knowledge Management}. Toronto, Canada, 539--548.

\bibitem[\protect\citeauthoryear{Wetherill and Glazebrook}{Wetherill and
  Glazebrook}{1986}]{wg86:seqmeth}
{\sc Wetherill, G.~B.} {\sc and} {\sc Glazebrook, K.~D.} 1986.
\newblock {\em Sequential Methods in Statistics\/} 3rd Ed.
\newblock Chapman and Hall.

\bibitem[\protect\citeauthoryear{Zobel}{Zobel}{1998}]{z98:sigir}
{\sc Zobel, J.} 1998.
\newblock How reliable are the results of large-scale information retrieval
  experiments?
\newblock In {\em Proc. 21st Annual International {ACM} {SIGIR} Conference on
  Research and Development in Information Retrieval}, {W.~B. Croft},
  {A.~Moffat}, {C.~J. van Rijsbergen}, {R.~Wilkinson}, {and} {J.~Zobel}, Eds.
  Melbourne, Australia, 307--314.

\end{thebibliography}

\received{May 2012}{August 2012}{October 2012}

\end{document}